\documentclass[letterpaper,twocolumn,10pt]{article}
\usepackage{usenix-2020-09}

\usepackage[pdftex]{graphicx}
\usepackage{longtable}
\usepackage{amsfonts}
\usepackage{amsmath}
\usepackage{amsthm}
\usepackage{enumerate}
\usepackage{pifont}
\usepackage{hyperref}
\usepackage{bm}
\usepackage{algorithmicx}
\usepackage[noend]{algpseudocode}
\usepackage{algorithm}

\graphicspath{{img/}}
\newcounter{margin} % Counter for margin comments
\definecolor{scolor}{rgb}{1.0, 0.08, 0.58}
\definecolor{lcolor}{rgb}{0.0, 0.27, 0.13}
\definecolor{fcolor}{rgb}{0.1, 0.1, 0.44}
\definecolor{changecolor}{rgb}{1.0, 0, 0}
\definecolor{newcolor}{rgb}{0.4, 0, 0.8}
\definecolor{ucolor}{rgb}{0, .5, 0}

\DeclareMathOperator*{\argmin}{arg\,min}
\DeclareMathOperator{\Tr}{Tr}

\newcommand{\nkw}{n}
\newcommand{\ndocs}{{N_{d}}}
\newcommand{\ndocsaux}{N_{d}^{aux}}
\newcommand{\ntok}{m}
\newcommand{\nq}{\rho}

\newcommand{\kw}[1]{{k_{#1}}}
\newcommand{\tok}[1]{{\tau_{#1}}}
\newcommand{\kwu}{\Delta_\kw{}}
\newcommand{\toku}{\Delta_\tok{}}
\newcommand{\kwufree}{\kwu^\circ}
\newcommand{\tokufree}{\toku^\circ}
\newcommand{\kwufixed}{\kwu^\bullet}
\newcommand{\tokufixed}{\toku^\bullet}
\newcommand{\obs}{\texttt{obs}}
\newcommand{\aux}{\texttt{aux}}
\newcommand{\ap}[1]{\mathbf{a}_{#1}}
\newcommand{\rep}[1]{r_{#1}}
\newcommand{\nrep}{2\nkw}

\newcommand{\Vobs}{\mathbf{V}}
\newcommand{\Vexp}{\hat{\mathbf{V}}}
\newcommand{\Vaux}{\tilde{\mathbf{V}}}
\newcommand{\Fobs}{\mathbf{F}}
\newcommand{\Faux}{\tilde{\mathbf{F}}}
\newcommand{\fobs}{\mathbf{f}}
\newcommand{\faux}{\tilde{\mathbf{f}}}
\newcommand{\freal}{\mathbf{f}^{real}}
\newcommand{\fdum}{\mathbf{f}^{dum}}
\newcommand{\fdumaux}{\tilde{\mathbf{f}}^{dum}}
\newcommand{\Freal}{{\mathbf{F}^{real}}}
\newcommand{\Fexp}{{\hat{\mathbf{F}}}}

\newcommand{\Fold}{{\mathbf{F}_\textsc{jan}^\textsc{jun}}}
\newcommand{\Fnew}{{\mathbf{F}_\textsc{jul}^\textsc{dec}}}
\newcommand{\Vauxnot}{\tilde{\mathbf{V}}^{\textsc{not}}}
\newcommand{\reptokw}{\nu}

\renewcommand{\P}{\Perm}
\newcommand{\Pfree}{\P^\circ}
\newcommand{\Pfixed}{\P^\bullet}
\newcommand{\Perm}{\mathbf{P}}
\newcommand{\perm}[1]{p(#1)}
\newcommand{\LAP}{\ensuremath{SolveLinear}}

\newcommand{\niters}{n_{iters}}
\newcommand{\pfree}{p_{free}}

\newcommand{\sap}{\texttt{SAP}}
\newcommand{\ikk}{\texttt{IKK}}
\newcommand{\graphm}{\texttt{GraphM}}
\newcommand{\fgraphm}{\texttt{FastPFP}}
\newcommand{\umeyama}{\texttt{Umeyama}}

\newcommand{\freq}{\texttt{Freq}}

\newcommand{\name}{\texttt{IHOP}}

\newcommand{\pancake}{\textsc{pancake}}

\newcommand{\osse}{\text{OSSE}}
\newcommand{\clrz}{\text{CLRZ}}
\newcommand{\TPR}{\text{TPR}}
\newcommand{\FPR}{\text{FPR}}

\usepackage{multirow}

\begin{document}

\date{}
\author{
{\rm Simon Oya}\\
University of Waterloo
\and
{\rm Florian Kerschbaum}\\
University of Waterloo
} % end author
\title{\Large \bf IHOP: Improved Statistical Query Recovery against\\Searchable Symmetric Encryption through Quadratic Optimization}
\maketitle
{\let\thefootnote\relax\footnotetext{\noindent{\textit{To appear in:} Proceedings of the 31st USENIX Security Symposium August 10–12, 2022, Boston, MA, USA\\
\url{https://www.usenix.org/conference/usenixsecurity22}}}}

% Concepts for title ideas: statistical-based attacks, ground-truth is not needed, statistical query recovery, IHOP
% IHOP: Improving Statistical Query Recovery against SSE Schemes through Quadratic Optimization

\begin{abstract}
Effective query recovery attacks against Searchable Symmetric Encryption (SSE) schemes typically rely on auxiliary \emph{ground-truth} information about the queries or dataset.
Query recovery is also possible under the weaker \emph{statistical} auxiliary information assumption, although statistical-based attacks achieve lower accuracy and are not considered a serious threat.
In this work we present $\name$, a statistical-based query recovery attack that formulates query recovery as a quadratic optimization problem and reaches a solution by iterating over linear assignment problems.
We perform an extensive evaluation with five real datasets, and show that $\name$ outperforms all other statistical-based query recovery attacks under different parameter and leakage configurations, including the case where the client uses some access-pattern obfuscation defenses.
In some cases, our attack achieves almost perfect query recovery accuracy.
Finally, we use $\name$ in a frequency-only leakage setting where the client's queries are correlated, and show that our attack can exploit query dependencies even when $\pancake$, a recent frequency-hiding defense by Grubbs et al., is applied.
Our findings indicate that statistical query recovery attacks pose a severe threat to privacy-preserving SSE schemes.
\end{abstract}

\section{Introduction}

%The increasing popularity of cloud-storage services in the recent years has brought new privacy concerns when the service provider is untrusted.
%While encrypting the dataset before outsourcing it preserves the data confidentiality, naively doing so deprives the client from certain functionalities such as database search.
Searchable Symmetric Encryption (SSE) schemes~\cite{song2000practical} allow a client to securely outsource a dataset to a cloud-storage provider, while still being able to perform secure queries over the dataset.
Efficient SSE schemes~\cite{curtmola2011searchable, chang2005privacy, kamara2012dynamic, cash2013highly, lau2014mimesis, kurosawa2014garbled, ogata2013toward, naveed2014dynamic, stefanov2014practical, he2014shadowcrypt, bost2016ovarphiovarsigma} typically leak certain information during their initialization step or querying process, that an honest-but-curious service provider could exploit to recover the database or guess the client's queries.
This leakage typically consists of the access pattern, which refers to the identifiers of the documents that match a query, and the search pattern, which refers to whether or not two queries are identical.

%Since their inception~\cite{song2000practical, curtmola2011searchable}, many authors have proposed different SSE schemes that vary in the type of queries they admit, their privacy properties, and their bandwidth, computation, and storage requirements.\smargin{add citations}
%
%One popular type of query that some SSE schemes provide is the \emph{point query}.
%In this case, the database consists of a set of documents, each labeled with at least one \emph{keyword}, and the user wants to be able to query for all the documents that match a particular keyword.
%Leakage-free SSE schemes that provide are unfeasible in terms of computation or bandwidth costs.
%Feasible schemes, however, leak information during their initialization step or the querying process, that an honest-but-curious service provider could exploit to recover the database or guess the client's queries.
%Cash et al.~\cite{cash2015leakage} define different leakage profiles L1-L4 that range from high leakage (L4) to low leakage (L1).
%This leakage typically consists of the access pattern, which refers to the identifiers of the documents that match a query, and the search pattern, which refers to whether or not two queries were generated with the same keyword.

There are many attacks that exploit access and search pattern leakage, as well as auxiliary information, to either recover the underlying keywords of the client's queries, or the dataset itself.
In this work, we study \emph{passive query recovery} attacks~\cite{islam2012access, liu2014search, cash2015leakage, pouliot2016shadow, blackstone2020revisiting, oya2021hiding, damie2021highly, ning2021leap} against SSE schemes that provide \emph{keyword query} functionality.
Namely, the dataset is a set of documents, each labeled with a list of keywords, and the client wants to be able to query for all the documents that match a particular keyword.
The adversary is an honest-but-curious service provider that follows the SSE protocol, but might want to infer the keywords of the client's queries from passively observing the system.
We can broadly classify passive query recovery attacks depending on the nature of their auxiliary information into attacks that assume partial ground-truth knowledge about the dataset and/or the underlying keywords of the client's queries (\emph{ground-truth attacks})~\cite{islam2012access, cash2015leakage, blackstone2020revisiting, damie2021highly, ning2021leap}, and those that \emph{only} require statistical information about the keyword distribution in the dataset and the client's querying behavior (\emph{statistical-based attacks})~\cite{islam2012access, pouliot2016shadow, oya2021hiding}.
Previous work has shown that ground-truth attacks can achieve high query recovery rates~\cite{blackstone2020revisiting, damie2021highly, ning2021leap}, and thus pose a significant privacy threat to SSE schemes.
Statistical-based attacks achieve lower query recovery accuracy, since they rely on a weaker auxiliary information assumption, although they are easier to adapt against access and search-pattern defense techniques~\cite{oya2021hiding}.
%Attacks that rely on statistical information can be adapted against padding defenses, but they typically incur high computational costs and, since their auxiliary information is only statistical, do not achieve high query recovery rates.
%Some attacks that can be used with statistical information are only truly effective when this information has been computed from the ground-truth dataset~\cite{islam2012access, cash2015leakage}.

In this work we focus on statistical-based query recovery attacks.
The strongest attacks from this family are $\graphm$ by Pouliot and Wright~\cite{pouliot2016shadow} and $\sap$ by Oya and Kerschbaum~\cite{oya2021hiding}.
$\graphm$ performs query recovery by heuristically solving a quadratic optimization problem.
This attack exploits the so-called volume co-occurrence information, which refers to how many documents two queries have in common, but uses the expensive PATH algorithm~\cite{zaslavskiy2008path}, and thus only works when the keyword universe size is small.
On the other hand, $\sap$~\cite{oya2021hiding} efficiently solves a linear problem using individual keyword volume information as well as query frequency information, but cannot take into account volume co-occurrence. % due to the linear nature of their attack.
In this paper, we propose a new statistical-based query recovery attack, that we call $\name$ since it follows an Iterative Heuristic algorithm to solve a quadratic Optimization Problem.
$\name$ can be used with any quadratic objective function, and thus it can combine volume and frequency leakage information like $\sap$~\cite{oya2021hiding}, exploit volume co-occurrence terms like $\graphm$~\cite{pouliot2016shadow}, and is the first attack to exploit frequency co-occurrence terms.
We use $\name$ to optimize a maximum-likelihood-based objective function, and evaluate our attack in five real datasets under different settings.
First, we consider the case where the adversary only uses volume leakage, where $\graphm$ is state-of-the-art.
We show that, both in SSE schemes that fully leak the access patterns during initialization~\cite{he2014shadowcrypt, lau2014mimesis} and in those that (at least) leak the access pattern of those keywords that are queried~\cite{curtmola2011searchable, chang2005privacy, kamara2012dynamic, cash2013highly, kurosawa2014garbled, ogata2013toward, naveed2014dynamic, stefanov2014practical, bost2016ovarphiovarsigma}, $\name$ consistently outperforms $\graphm$, achieving higher accuracy while being orders of magnitude faster (e.g., $\name$ achieves $\approx99\%$ accuracy on Lucene dataset with $\nkw=1\,000$ keywords and finishes in 15 minutes, while $\graphm$ is achieves $\approx56\%$ accuracy and takes almost three days to finish).
Second, we consider the case where the adversary uses frequency leakage and the client sends queries independently, where $\sap$ is state-of-the-art.
We show that $\name$ comfortable outperforms $\sap$ and all other attacks, even in the case where the client uses two access-pattern-hiding defenses~\cite{chen2018differentially, shang2021obfuscated}.

Finally, we show that $\name$ can also exploit query correlations in query recovery.
For this, we consider a case where the client queries dataset entries individually and thus the attack must rely on query frequency information only.
$\pancake$, a system proposed by Grubbs et al.~\cite{grubbs2020pancake}, provably ensures that the frequency of access to each dataset entry is the same.
Although $\pancake$ does not hide query correlations, a preliminary analysis shows that it still provides sufficient protection in that case~\cite{cryptoeprint:2020:1501}.
We adapt $\name$ against $\pancake$ when client's queries follow a Markov process, and show that our attack can effectively exploit query correlations to recover keywords (in our example, $\name$ achieves $\approx48\%$ accuracy when the adversary has low-quality auxiliary information about the client's querying behavior, and $\approx69\%$ accuracy when the adversary has high-quality information).

In summary, we propose a statistical-based query recovery attack that outperforms all other known attacks when the adversary does not have ground-truth information on the client's dataset or queries~\cite{islam2012access, liu2014search, pouliot2016shadow, oya2021hiding}.
Our results show that statistical-based query recovery attacks pose a serious threat to SSE schemes, since they can achieve high recovery rates without access to ground-truth information and can adapt against volume and frequency-hiding defenses.

\section{Related work} 
\label{sec:relwork}

%\sinline{From Hahn: Song et al.~\cite{song2000practical} are the first? Then Curtmola et al.~\cite{curtmola2011searchable} present the first SSE construction (using inverted index).
%Then Blackstone et al.~\cite{blackstone2020revisiting} divides attack into those that require full or partial knowledge of the database (known-data) and those that require similar data. They mention that IKK was the first, then Cash, Blackstone, Ning~\cite{ning2018passive} and Poddar~\cite{poddar2020practical}. They mention most of these attacks can be executed as similar-data attacks but are only effective as known-data attacks.
%
%There is a line for database reconstructions with range queries~\cite{grubbs2019learning, lacharite2018improved}}

%Attacks against SSE schemes vary depending on their goal, adversary model, targeted leakage profile, and their required auxiliary information, as summarize by Blackstone et al.~\cite{blackstone2020revisiting}.
%The focus of this work are those attacks whose goal is to recover the underlying keywords of the client's queries (\emph{query recovery attacks}).
%Depending on their goal, attacks can aim at recovering the client's queries~\cite{} or the database itself.
%Database recovery attacks are typically aimed against 
%Attacks against strong schemes that only leak the number of documents returned (but not \emph{which} documents) are typically based on expensive primitives like ORAM~\cite{goldreich1996software} and PIR~\cite{}
%
%Searchable Symmetric Encryption (SSE)~\cite{song2000practical, curtmola2011searchable} is a popular primitive

Since their inception~\cite{song2000practical, curtmola2011searchable}, many authors have proposed SSE schemes with different privacy and utility trade-offs.
Designing SSE schemes without access pattern leakage implies leveraging expensive primitives such as ORAM~\cite{goldreich1996software} or PIR~\cite{chor1995private}, which incurs expensive bandwidth, computational, and storage costs, while still being vulnerable to certain volume-based attacks~\cite{grubbs2018pump, kellaris2016generic, poddar2020practical}.
In this work we target \emph{efficient and deployable SSE} schemes that leak the access and search patterns~\cite{song2000practical, curtmola2011searchable, chang2005privacy, kamara2012dynamic, cash2013highly, lau2014mimesis, kurosawa2014garbled, ogata2013toward, naveed2014dynamic, stefanov2014practical, he2014shadowcrypt, bost2016ovarphiovarsigma}.
%Cash et al.~\cite{cash2015leakage} provide a classification of typical leakage profiles in these schemes, that we refer to below.
We consider SSE schemes where the client performs \emph{keyword queries} (i.e., the client queries for a single keyword to retrieve all documents that match that keyword) and the adversary is an honest-but-curious service provider that performs a \emph{query recovery} attack to guess the underlying keywords of the client's queries.

Query recovery attacks can be broadly classified depending on the auxiliary data they require into ground-truth attacks~\cite{islam2012access, cash2015leakage, blackstone2020revisiting, damie2021highly, ning2021leap}, which partially know some of the client's queries or database, and statistical-based attacks~\cite{islam2012access, liu2014search, pouliot2016shadow, oya2021hiding}, that have statistical information about the database and client's querying behavior (e.g., a set of non-indexed documents).

\paragraph{Ground-truth attacks}
Islam et al.~\cite{islam2012access} propose one of the first query recovery attacks ($\ikk$) that assumes full dataset knowledge and partial query knowledge. 
Their attack exploits access pattern leakage to compute volume co-occurrence matrices and solves a quadratic optimization problem using simulated annealing.
Cash et al.~\cite{cash2015leakage} propose the count attack, which improves upon $\ikk$ by requiring only partial database knowledge.
This attack follows an iterative algorithm that keeps reducing the set of candidate keywords for each unknown query until only one feasible candidate remains.
Blackstone et al.~\cite{blackstone2020revisiting} propose an attack based on subgraphs that follows different refinement heuristics to the count attack, outperforming it.
Recently, Damie et al.~\cite{damie2021highly} proposed a hybrid attack that uses ground-truth query knowledge, but only statistical dataset information, and Ning et al.~\cite{ning2021leap} propose a ground-truth attack that achieves both keyword and document recovery.

\paragraph{Statistical-based attacks}
Even though $\ikk$ originally assumes ground-truth dataset and query knowledge~\cite{islam2012access}, this attack can also be evaluated in the setting where only statistical information is available (see Sec.~\ref{sec:eval1}).
The graph matching attack ($\graphm$) by Pouliot and Wright~\cite{pouliot2016shadow} uses volume information computed from access-pattern leakage, like $\ikk$, but uses a more refined optimization function and looks for a solution using graph matching algorithms~\cite{umeyama1988eigendecomposition, zaslavskiy2008path}.
The frequency-only attack ($\freq$) by Liu et al.~\cite{liu2014search} exploits search-pattern leakage and auxiliary information about the client's querying behavior to perform query recovery.
Recently, Oya and Kerschbaum~\cite{oya2021hiding} proposed $\sap$, an attack that exploits both volume and frequency information to efficiently solve a linear assignment problem using known optimal solvers~\cite{fredman1987fibonacci}.

Even though query recovery rates achieved by statistical-based attacks are typically below those achieved by ground-truth attacks, due to their weaker auxiliary knowledge assumptions, statistical-based attacks can be easily tuned~\cite{oya2021hiding} against volume-padding defenses~\cite{chen2018differentially, demertzis2020seal, patel2019mitigating}.
Ground-truth attacks, however, perform poorly against such defenses~\cite{shang2021obfuscated}, since they are designed to use exact information about the data and the defenses typically randomize the leaked patterns.

\section{Problem Statement}
%
%We begin this section by providing a high-level description of our system model, and then we formalize our notation and the query recovery problem that we consider in this work.
%We then classify previous attacks that are relevant to our work according to how they solve the query recovery problem.

\subsection{Overview}

Our system model consists of two parties: a client and a server.
The client owns a privacy-sensitive dataset that she wishes to store remotely on the server.
The server offers storage services (e.g., it is a cloud storage provider), but is not trusted by the client.
The client encrypts each document using symmetric encryption, and sends the encrypted documents to the server.
Each document has a set of \emph{keywords} attached to it, and the client wishes to be able to issue \emph{keyword queries}, i.e., queries that retrieve all the documents that contain a particular a keyword.\footnote{Note that some works consider datasets where each document is associated to a \emph{single} keyword~\cite{demertzis2020seal, patel2019mitigating} (e.g., the keyword can be the document's publication date), and thus queries for different keywords always return disjoint sets of documents.
The attacks we consider in this paper rely on volume co-occurrence leakage, which only occurs when at least some documents have more than one keyword.}
In order to achieve this search functionality in the encrypted dataset, the client uses a Searchable Symmetric Encryption (SSE) scheme.
This SSE scheme has an initialization step where the client generates an \emph{encrypted search index}, which she sends to the server alongside the encrypted documents.
Later, when the client wishes to query for a particular keyword, she generates a \emph{query token} from that keyword and sends it to the server.
The server evaluates the query token in the encrypted search index, which reveals which encrypted documents should be returned to the client in response.

The server is honest-but-curious, i.e., it follows the protocol specifications, but might be interested in learning sensitive information from the client from passive observation.
We focus on an adversary that wants to guess the underlying keywords of the client's issued query tokens, i.e., to perform a \emph{query recovery attack}.
This can in turn help the adversary guess the keywords attached to each encrypted document (database recovery attack).

Even though encrypting the documents and hiding which documents contain each keyword through the encrypted search index prevents the adversary from trivially matching query tokens to keywords, most efficient SSE schemes leak certain information that can be used to carry out a query recovery attack, with the help of certain auxiliary information.
There are two typical sources of leakage in existing SSE schemes: access pattern leakage and search pattern leakage.

The access pattern of a query is the list of document identifiers that match the given query (i.e., the identifiers of the documents that the server returns to the client in response to the query).
The search pattern refers to whether or not two queries are identical (i.e., whether they have the same underlying keyword).
An adversary with \emph{auxiliary information} about how often the client queries for particular keywords, or how many documents have certain keywords, can leverage this leakage to carry out a query recovery attack. 

%\TODO{Summary of assumptions?
%\begin{enumerate}
	%\item Honest-but-curious adversary (no packet injection).
	%\item No ground-truth information about the dataset or the queries.
	%\item Auxiliary information is different from the ground truth.
	%\item The scheme leaks the access pattern, but the user can use obfuscation techniques (FN, FP). (either L1 or L2 leakage)
	%\item The scheme leaks the search pattern, but the user can use dummy queries and document replicas to hide query frequencies.
%\end{enumerate}
%}

\subsection{Formal Problem Description and Notation}

We formalize the problem described above and introduce our notation, which we summarize in Table~\ref{tab:notation}.
We note that this general description is not tailored to a specific SSE scheme, but accommodates a broad family of SSE schemes that leak the access and search patterns~\cite{song2000practical, curtmola2011searchable, chang2005privacy, kamara2012dynamic, cash2013highly, lau2014mimesis, naveed2014dynamic, stefanov2014practical, he2014shadowcrypt, bost2016ovarphiovarsigma}.

\begin{table}[t]
\centering
\begin{tabular}{ r  l }
%\hline
%\multicolumn{2}{c}{General Parameters} \\ \hline
%\textbf{Symbol} & \textbf{Meaning} \\ \hline
$\ndocs$ & Number of documents in the dataset.\\
$\nkw$ & Total number of keywords, $\nkw\doteq|\kwu|$.\\
$\ntok$ & Total number of observed query tokens, $\ntok\doteq|\toku|$.\\
$\nq$ & Number of queries issued by the client.\\
$D$ & Dataset $D\doteq\{d_1,d_2,\dots,d_\ndocs\}$ \\
$\kwu$ & Keyword universe $\kwu\doteq[\kw{1},\kw{2},\dots,\kw{\nkw}]$. \\
$\toku$ & Token universe $\toku\doteq[\tok{1}, \tok{2}, \dots, \tok{\ntok}]$.\\
$\kw{i}$ & $i$th keyword, with $i\in[\nkw]$.\\
$\tok{j}$ & $j$th query token, with $j\in[\ntok]$.\\
$\ap{j}$ & Access pattern for token $\tok{j}$ ($\ndocs\times 1$).\\ \hline
$\Vobs$ & Matrix of observed token volumes ($\ntok\times\ntok$).\\
$\Vaux$ & Matrix of auxiliary keyword volumes ($\nkw\times\nkw$).\\
$\fobs$ & Vector of observed token frequencies ($\ntok\times 1$).\\
$\faux$ & Vector of auxiliary keyword frequencies ($\nkw\times 1$).\\
$\Fobs$ & Markov matrix of observed token freqs.~($\ntok\times\ntok$).\\
$\Faux$ & Markov matrix of auxiliary keyword freqs.~($\nkw\times\nkw$).\\
\end{tabular}
\caption{Summary of notation \label{tab:notation}}
\end{table}

We use boldface lowercase characters for vectors (e.g., $\mathbf{a}$), and boldface uppercase characters for matrices (e.g., $\mathbf{A}$).
The transposition of $\mathbf{A}$ is $\mathbf{A}^T$, and $\mathbf{A}_{i,j}$ is the $i,j$th entry of $\mathbf{A}$.
All products between matrices and vectors are dot products.
We use $[n]\doteq\{1,2,\dots,n\}$ for a positive integer $n$.

We denote the client's dataset by $D\doteq\{d_1,d_2,\dots,d_\ndocs\}$, where $\ndocs$ is the number of documents.
We refer to the documents by their index, and assume that they are randomly shuffled during setup so that their index does not reveal any information about their content.
Each document has a set of keywords attached to it, and keywords belong to the \emph{keyword universe} $\kwu\doteq\{\kw{1}, \kw{2}, \dots, \kw{\nkw}\}$, of size $\nkw$.
The client encrypts each document $d\in D$, uses an SSE scheme to generate an encrypted search index, %$\Ienc$,\smargin{we don't use it, I might remove it} 
and sends the encrypted dataset and index to the server.
In order to query for a particular keyword $\kw{}\in\kwu$, the client first generates a query token $\tok{}$ using $\kw{}$, and she sends it to the server.
The server evaluates the query token $\tok{}$ on the encrypted search index, which reveals the access pattern, i.e., the indices of the documents that match the query.
We represent the access pattern of a token $\tok{j}$ as a column vector $\ap{j}$ of length $\ndocs$ whose $\ell$th entry is 1 if $d_\ell$ matches the query, and 0 otherwise.
We consider SSE schemes that reveal the search pattern, i.e., they reveal whether or not two query tokens were generated with the same underlying keyword.
Thus, in our notation we use $\tok{j}$ and $\tok{j'}$ ($j\neq j'$) for two tokens that have been generated with different keywords.
We use $\toku\doteq\{\tok{1},\tok{2},\dots,\tok{\ntok}\}$ to denote the set of all ($\ntok$) distinct tokens observed by the adversary.

\subsubsection{Adversary's observation and auxiliary information}\label{sec:obs}
We use $\obs$ to denote the adversary's observation, which corresponds to both the leakage of the SSE scheme during its initialization step and while the client performs queries.
Typically, $\obs$ comprises the access and search patterns of the queries, i.e., $\obs=[(\tok{},\ap{})_1, (\tok{},\ap{})_2, \dots, (\tok{},\ap{})_\nq]$, where $(\tok{},\ap{})_r$ is the query token and access pattern of the $r$th query.
Most statistical-based query recovery attacks compute certain summary statistics from the observation before running the attack~\cite{islam2012access, cash2015leakage, pouliot2016shadow, oya2021hiding}.
These statistics are typically the \emph{volume} and \emph{frequency} information, which we define next.
The matrix of observed volumes $\Vobs$ is an $\ntok\times\ntok$ matrix whose $j,j'$th entry contains the number of documents that match both query tokens $\tok{j}$ and $\tok{j'}$, i.e., $\Vobs_{j,j'}=\ap{j}^T \ap{j'} /\ndocs$.
The vector of observed query frequencies $\fobs$ is a vector of length $\ntok$ whose $j$th entry contains the number of times the client queried for $\tok{j}$, normalized by the total number of queries $\nq$.
The (Markov) matrix of observed query frequencies $\Fobs$ is an $\ntok\times\ntok$ matrix whose $j,j'$th entry contains the number of times the client sent token $\tok{j'}$ followed by $\tok{j}$, normalized by the total number of queries with token $\tok{j'}$ (which we denote by $\nq(\tok{j'}))$.

We use $\aux$ to denote the adversary auxiliary's information.
Since we consider statistical-based attacks, this information does not contain ground-truth knowledge about the queries and/or dataset.
The adversary uses $\aux$ to compute the vectors and matrices $\Vaux$, $\faux$, $\Faux$, whose structure is the same as the variables that the adversary can compute for the observations ($\Vobs$, $\fobs$, $\Fobs$), defined above.
In this paper, we generate the auxiliary information related to keyword volume by giving the adversary non-indexed documents~\cite{pouliot2016shadow, oya2021hiding, damie2021highly} (i.e., documents not in the client's dataset), and auxiliary frequency information by giving the adversary outdated query frequencies~\cite{oya2021hiding}.
Note that the observed variables ($\Vobs$, $\fobs$, $\Fobs$) refer to volume and frequency statistics computed from the \emph{query tokens}, but the auxiliary variables ($\Vaux$, $\faux$, $\Faux$) refer to the \emph{keywords}.
For example, $\Vaux$ is an $\nkw\times\nkw$ matrix whose $i,i'$th entry contains an \emph{approximation} of the percentage of documents that have both keywords $\kw{i}$ and $\kw{i'}$ in the dataset.
In most of the SSE schemes we consider in this paper, there is a one-to-one correspondence between keywords and tokens, which the adversary aims to guess.\footnote{We also consider two schemes where different tokens might correspond to the same keyword~\cite{shang2021obfuscated, grubbs2020pancake}; we explain the details in Sections~\ref{sec:freq} and \ref{sec:eval2}.}
We note that auxiliary information is \emph{imprecise}: the actual percentage of documents with keywords $\kw{i}$ and $\kw{i'}$ in the dataset will likely differ from $\Vaux_{i,i'}$.

\subsubsection{Leakage scenarios}
We consider three different leakage scenarios in our paper. 
These scenarios abstract from the actual SSE scheme being used, and all SSE schemes that have \emph{at least} this leakage are subject to the attacks we study in that setting.
\begin{itemize}
	\item \textbf{S1: full access-pattern leakage, no queries.} 
	In this scenario, the client uses a scheme whose initialization step leaks which documents are returned as a response to each query token.
	(This corresponds to the leakage type $L2$ by Cash et al.~\cite{cash2015leakage}.)
	A simple example of this is a scheme where the client uses deterministic encryption to generate a query token from a keyword, and simply sends the server the list of encrypted documents with the query tokens corresponding to the keywords they contain.
	In this setting, we run the attack before the client has performed any query.
	The adversary's observation is therefore $\obs=[\ap{1},\ap{2},\dots,\ap{\ntok}]$, and thus the attack relies solely on volume information ($\Vobs$ and $\Vaux$).
	
	This setting includes academic proposals~\cite{he2014shadowcrypt, lau2014mimesis} as well as commercial products (e.g., see Cash et al.~\cite{cash2015leakage}).
	
	\item \textbf{S2: access-pattern leakage for queries.}
	In this scenario, the initialization step does not leak any information.
	The client issues queries, and each token $\tok{}$ leaks its access pattern $\ap{}$, i.e, the list of documents returned in response to that token.
		(This corresponds to the leakage type $L1$ by Cash et al.~\cite{cash2015leakage}.)
	Let $(\tok{},\ap{})_r$ be the query token and access pattern of the $r$th query and assume that the client performs $\nq$ queries in total.
	The adversary's observation is $\obs=[(\tok{},\ap{})_1, (\tok{},\ap{})_2, \dots, (\tok{},\ap{})_\nq]$
	
	In this scenario, some keywords might never be queried, so the adversary might not see all possible access patterns ($\ntok\leq\nkw$), contrary to $S1$.
	We study two cases: when the adversary does not have auxiliary frequency information and thus it relies on volume information only ($\Vobs$ and $\Vaux$), and when the adversary can additionally exploit independent query frequencies ($\fobs$ and $\faux$) to aid the attack.
	
	Most efficient SSE schemes exhibit at least this leakage~\cite{song2000practical, curtmola2011searchable, chang2005privacy, kamara2012dynamic, cash2013highly, lau2014mimesis, kurosawa2014garbled, ogata2013toward, naveed2014dynamic, stefanov2014practical, he2014shadowcrypt, bost2016ovarphiovarsigma}
	
	\item \textbf{S3: frequency-only leakage.}
	Finally, we consider a setting identical to $S2$ where each keyword matches exactly one document, and no two keywords match the same document.
	This means that $\nkw=\ndocs$ and, without loss of generality, $\kw{i}$ matches only $d_i$.
	In this case, only frequency information is useful for query recovery: $\obs=[(\tok{})_1, (\tok{})_2, \dots, (\tok{})_\nq]$.
	
	We consider this setting to evaluate the effectiveness of our attack in the presence of query dependencies without volume leakage (i.e., the adversary only uses $\Fobs$ and $\Faux$).
	This is important since our attack is the first query recovery attack that can exploit query correlations (besides the attack example in the appendix of~\cite{cryptoeprint:2020:1501}).	
\end{itemize}

\subsubsection{Adversary's goal and success metrics}
The goal of the adversary is to carry out a query recovery attack, i.e., to find the underlying keywords of each query token.
The outcome of the attack is therefore an injective (one-to-one) mapping from the set of query tokens to the set of keywords, which we denote by $\perm{\cdot}: [\ntok]\to[\nkw]$.
For example, $\perm{j}=i$ denotes that the adversary guesses that the underlying keyword of token $\tok{j}$ is $\kw{i}$.
We sometimes represent this mapping as an $\nkw\times\ntok$ matrix $\P$ where $\P_{i,j}=1$ if $\perm{j}=i$, and 0 otherwise. %whose $i,j$th component is  :
%\begin{equation*}
 %\P_{i,j}=\begin{cases}
	%1\,, &\text{if }\perm{j}=i\,,\\
	%0\,, &\text{otherwise}\,.
	%\end{cases}
%\end{equation*}
We use $\mathcal{P}$ to denote the set of all valid mappings $\P$, i.e., $\mathcal{P}\doteq\{\P| \mathbf{1}_\nkw^T\P=\mathbf{1}^T_\ntok, \P\mathbf{1}_{\ntok}\leq\mathbf{1}_\nkw, \P\in\{0, 1\}^{\nkw\times\ntok}\}$, where $\mathbf{1}_\nkw$ is an all-ones column vector of length $\nkw$.

A query recovery attack takes the observations $\obs$ and auxiliary information $\aux$ as input, and produces a mapping of query tokens to keywords $\P$ as output.
We measure the success of an attack as the \emph{percentage of observed query tokens} ($\tok{}\in\toku$) for which the attack \emph{correctly} guesses their underlying keyword.
This is the most popular success metric in previous works, and it is referred to as the \emph{attack accuracy}~\cite{islam2012access, liu2014search, pouliot2016shadow, oya2021hiding} or the query \emph{recovery rate}~\cite{cash2015leakage, blackstone2020revisiting}. %\smargin{I've noticed I used ``query recovery rate'' before this because it's more ``descriptive'', but attack accuracy later. Should I just use one? (in that case I think I prefer the recovery rate?)}
%In Appendix~\ref{app:extra_results} we also include results for the \emph{weighted accuracy}, which is the \emph{percentage of queries} for which the attack correctly guesses their underlying keyword.
%The weighted accuracy gives a higher \emph{weight} to those tokens that are queried more often.
%
%Throughout the paper, we us
%Let $\P^*$ be the \emph{true} mapping of tokens to keywords.
%The \emph{unweighted accuracy} $\accu$ is the number of tokens $\tok\in\toku$ for which the mapping $\P$ correctly identifies their keyword, divided by the total number of tokens, i.e.,
%\begin{equation}
	%\accu(\P)\doteq\frac{1}{\ntok}\sum_{j=1}^\ntok I(\perm{j}=\perm{j}^*)\,,
%\end{equation}
%where $I(\cdot)$ is the indicator function ($I(\cdot)=1$ when its argument is true, otherwise $I(\cdot)=0$).
%The \emph{weighted accuracy} $\accw$ is the number of queries for which the mapping $\P$ correctly identifies their keyword, divided by the total number of queries.
%If $\nq(\tok{j})$ is the number of queries the client did sending token $\tok{j}$, then
%\begin{equation}
	%\accw(\P)\doteq\frac{1}{\nq}\sum_{j=1}^\ntok \nq(\tok{j}) I(\perm{j}=\perm{j}^*)\,.
%\end{equation}
%As their name suggests, the unweighted accuracy measures the attack performance without giving more importance to popular queries, while the weighted accuracy is more affected by the accuracy in identifying popular queries, and less by rare queries.

\section{Quadratic Query Recovery with $\name$}

%\smargin{Make the attack description less technical, easier to follow. More accessible, maybe give a higher level description.}
In this section, we present our query recovery attack, which we call $\name$ (Iteration Heuristic for quadratic Optimization Problems).
We begin with an overview of statistical query recovery attacks, noticing that some solve linear optimization problems, while others are based on quadratic problems.
We then propose $\name$, which uses a linear optimization solver to iteratively look for a (suboptimal) solution to a quadratic query recovery problem.
$\name$ is a general algorithm that can be used to minimize different quadratic objective functions.
We particularize it to use the volume and frequency statistics we mentioned in Section~\ref{sec:obs}, using a maximum likelihood-based objective function.
We evaluate the performance of $\name$ in Sections~\ref{sec:eval1} and \ref{sec:eval2}.

\subsection{Linear and Quadratic Query Recovery}

Most query recovery attacks can be framed as an optimization problem, where the adversary tries to find the assignment of keywords to query tokens $\P$ that minimizes a certain objective function.
These problems are typically linear or quadratic with respect to $\P$.
Linear query recovery attacks~\cite{liu2014search,oya2021hiding} can be formulated as
\begin{equation} \label{eq:lap}
	\P=\argmin_{\P\in\mathcal{P}} \sum_{i\in[\nkw]} \sum_{j\in[\ntok]} c_{i,j,} \cdot \P_{i,j} \,.
\end{equation}
This problem follows the structure of a Linear Assignment Problem (LAP), which can be optimally solved with a computational cost of $O(\nkw\cdot\ntok + \ntok^2\cdot\log\ntok)$~\cite{fredman1987fibonacci}.
The constants $c_{i,j}$ represent the cost of assigning keyword $\kw{i}$ to token $\tok{j}$.

Quadratic query recovery attacks follow the formulation
\begin{equation} \label{eq:qap}
	\P=\argmin_{\P\in\mathcal{P}} \sum_{i,i'\in[\nkw]} \sum_{j,j'\in[\ntok]} c_{i,i',j,j'} \cdot \P_{i,j} \cdot \P_{i',j'}\,,
\end{equation}
where $c_{i,i',j,j'}$ is the cost of jointly assigning keyword $\kw{i}$ to token $\tok{j}$ and $\kw{i'}$ to token $\tok{j'}$.
This mathematical problem follows Lawler's~\cite{lawler1963quadratic} formulation of the general Quadratic Assignment Problem (QAP), which is known to be NP-complete~\cite{sahni1976p}.
Existing attacks that follow this formulation thus rely on suboptimal heuristic algorithms to find a solution~\cite{islam2012access, pouliot2016shadow}.

\subsection{Quadratic Query Recovery via Linear Assignments}

While there are efficient optimal solvers for the LAP \eqref{eq:lap}, solvers for the QAP are suboptimal and heuristic.
%The LAP however cannot take into account
%As mentioned above, there exist efficient solvers for the LAP \eqref{eq:lap}, but not for the QAP \eqref{eq:qap}.
However, attacks based on a LAP cannot exploit the quadratic terms in a QAP that can contain valuable information for query recovery.
We present $\name$, a query recovery attack that relies on efficient solvers for the LAP to iteratively solve a QAP.
%Our attack is not optimal (it runs in polynomial time, so it cannot find the optimal solution to a QAP, which is NP-complete), but it is intuitively more suitable for query recovery than other QAP solvers, as we verify later in our evaluation.
%\smargin{Sketch of the Hungarian algorithm}

\begin{figure}
\begin{center}
\def\svgwidth{\linewidth} 
{
%\fontsize{7.5pt}{9.5pt}
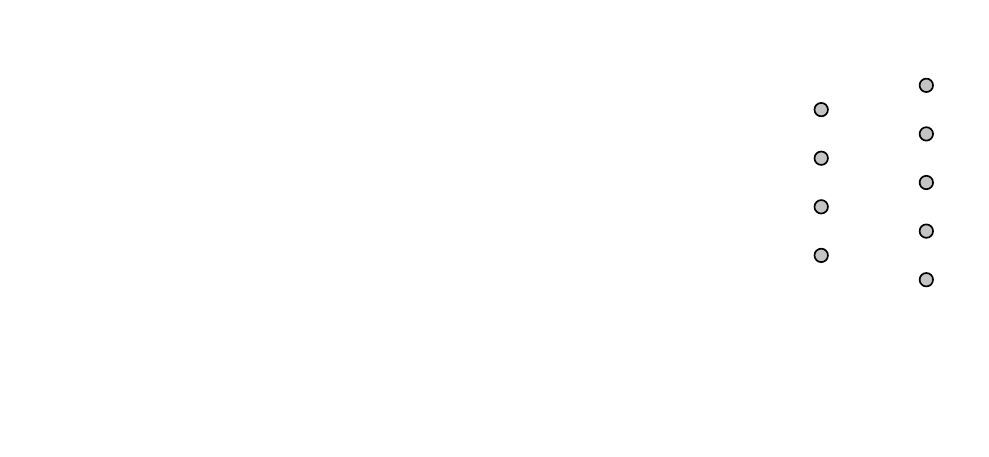
}
\caption{One iteration of $\name$ \label{fig:attack}}
\end{center}
\end{figure}

We first explain a single iteration of our attack at a high level, and then we present its formal description.
Figure~\ref{fig:attack} shows a toy example of an iteration of our attack.
At the beginning of this iteration, the attack has an assignment of tokens to keywords $\P$ (Fig.~\ref{fig:attack}a).
Then, it \emph{fixes} some of those assignments at random (denoted $\Pfixed$, see Fig.~\ref{fig:attack}b) and \emph{frees} the remaining keywords and query tokens.
The sets of free query tokens $\tokufree$, fixed query tokens $\tokufixed$, free keywords $\kwufree$, and fixed keywords $\kwufixed$ for this example are shown in the bottom of the figure.
The attack re-computes an assignment of the free query tokens to the free keywords, leveraging the fixed assignment $\Pfixed$ to improve the re-estimation (Fig.~\ref{fig:attack}c).
This yields an update of the assignment $\P$ and ends the iteration (Fig.~\ref{fig:attack}d).
The idea of this approach is that, if some assignments are fixed, we can use the quadratic terms involving those assignments and the assignments we wish to update while keeping the optimization problem linear. 

\begin{algorithm}[t]
    \caption{$\name$}
    \label{alg:attack}
    \begin{algorithmic}[1] % The number tells where the line numbering should start
        \Procedure{$\name$}{$\aux,\obs,\niters,\pfree$} %\Comment{The g.c.d. of a and b}
            \State $\P\gets \LAP(\kwu,\toku,\emptyset,\emptyset,\emptyset,\aux,\obs)$ \label{line:init}
            \For{$i\gets 1 \dots \niters$} \label{line:iter}
                \State $\tokufree\xleftarrow{\lceil\pfree\cdot\ntok\rceil}\toku$ \label{line:tokufree} \Comment{Choose $\pfree\cdot\ntok$ tokens}
                \State $\tokufixed = \{\tok{j} | \tok{j}\in\toku, \tok{j}\notin\tokufree\}$ \label{line:tokufixed} \Comment{Fixed tokens}
								\State $\kwufixed = \{\kw{i}|i=p(j), \kw{i}\in\kwu, \tok{j}\in\tokufixed\}$ \label{line:kwufixed}
								\State $\kwufree = \{\kw{i}|\kw{i}\in\kwu, \kw{i}\notin\kwufixed\}$ \label{line:kwufree}
								\State $\Pfixed = \{\tok{j}\to\kw{p(j)}| \tok{j} \in \tokufixed\}$ \label{line:Pfixed}
                \State $\Pfree\gets \LAP(\kwufree, \tokufree, \kwufixed, \tokufixed, \Pfixed, \aux, \obs)$ \label{line:LAP}
								\State $\P\gets\text{combine}(\Pfree,\Pfixed)$ \label{line:enditer}
            \EndFor
            \State \textbf{return} $\P$
        \EndProcedure
    \end{algorithmic}
\end{algorithm}

\begin{algorithm}[t]
    \caption{$\LAP$}
    \label{alg:LAP}
    \begin{algorithmic}[1] % The number tells where the line numbering should start
        \Procedure{$\LAP$}{$\kwufree, \tokufree, \kwufixed, \tokufixed, \Pfixed, \aux, \obs$} 
            \State Get $c\equiv\{c_{i,i',j,j'}\}$ and $d\equiv\{d_{i,j}\}$ using $\aux$, $\obs$.
            \State Solve the linear assignment problem:
\begin{equation*} \hspace{-15pt}
	\resizebox{\hsize}{!}{$
	\Pfree=\argmin\limits_{\Pfree\in\mathcal{P}^{\circ}}
	\displaystyle\sum\limits_{\kw{i}\in\kwufree} \sum\limits_{\tok{j}\in\tokufree} \left(
	\sum\limits_{\tok{j'}\in\tokufixed} \sum\limits_{\kw{i'}\in\kwufixed} c_{i,i',j,j'} \cdot \Pfree_{i,j} \cdot \Pfixed_{i',j'} + d_{i,j}\cdot \Pfree_{i,j}\right)\,.
	$}
\end{equation*} \label{line:opt}
				\State \textbf{return} $\Pfree$
        \EndProcedure
    \end{algorithmic}
\end{algorithm}

We formally describe the attack in Algorithm~\ref{alg:attack}.
The attack receives the observations $\obs$ and auxiliary information $\aux$, and two parameters: the number of iterations $\niters$ and the percentage of free tokens for each iteration $\pfree$.
The attack begins with an initialization step, where it computes $\P$ by solving a linear problem $\LAP$ that we specify below (Line~\ref{line:init}).
Then, it iterates $\niters$ times.
At each iteration, the attack splits the set of query tokens $\toku$ at random into two groups: the group of \emph{free} tokens $\tokufree$, which contains $\lceil\pfree\cdot\ntok\rceil$ tokens (Line~\ref{line:tokufree}), and the group of \emph{fixed} tokens $\tokufixed$, which contains all the other tokens (Line~\ref{line:tokufixed}). 
Let $\kwufixed$ be the set of keywords assigned to tokens in $\tokufixed$ by $\P$, and let $\kwufree$ be the set of all the keywords not in $\kwufixed$; i.e., the \emph{free} keywords (Lines \ref{line:kwufixed}--\ref{line:kwufree}).
The fixed matching $\Pfixed$ is the bijective mapping of $\tokufixed\to\kwufixed$ extracted from $\P$ (Line \ref{line:Pfixed}).
Then, the attack looks for an assignment $\Pfree$ of the free tokens $\tokufree$ to the free keywords $\kwufree$ by solving a linear problem (Line~\ref{line:LAP}).
The attack finally updates $\P$ using the newly computed $\Pfree$ and the fixed assignment $\Pfixed$, and this finishes the iteration.

A key component of this attack is the linear problem $\LAP$: this problem specifies how the adversary uses the auxiliary information $\aux$, the observations $\obs$, and the fixed mapping $\Pfixed$ to update the matching $\P$ in each iteration.
We show this problem in Algorithm~\ref{alg:LAP}.
This problem is an instantiation of a QAP \eqref{eq:qap} that only considers quadratic terms that depend on $\Pfree$ and $\Pfixed$, but never between two terms in $\Pfree$.
Thus, the problem is linear (in $\Pfree$) and can be written as a LAP~\eqref{eq:lap}. 
Since this problem is a LAP, it can be efficiently and optimally solved with the Hungarian algorithm~\cite{kuhn1955hungarian, lawler1963quadratic}.
The constants $c$ and $d$ are computed from $\obs$ and $\aux$, and their actual expression depends on the SSE setting in question; below, we give expressions for these constants for the volume and frequency leakage scenarios we defined in Section~\ref{sec:obs}.
The constants $d_{i,j}$ are simply the terms $c_{i,i,j,j}$ in \eqref{eq:qap}, that we write separately from $c$ in Algorithm~\ref{alg:LAP} to distinguish purely linear coefficients ($d$) from quadratic ones ($c$).
Our attack therefore provides a \emph{template} that researchers can use to instantiate quadratic query recovery attacks that optimize different objective functions, captured by $c$ and $d$.

\subsection{Coefficient Selection in $\name$}

We follow an approach inspired by maximum likelihood estimation to set the values of the coefficients ($c$ and $d$) in each iteration of $\LAP$.
In short, $\LAP$ finds the assignment between free keywords and tokens $\Pfree$ that maximizes the \emph{log-likelihood} of the observations $\obs$ given the auxiliary information $\aux$.
%We describe how we tune the coefficients below, and refer to the full version of the paper for a more technical explanation.

First, we present the expressions when only one source of leakage is available to the adversary: either volume leakage, frequency leakage with independent queries, or frequency leakage with correlated queries.
Then, we explain how we combine them when more than one source of leakage is available to the adversary.
To derive these expressions, we leverage \emph{binomial} and \emph{Poisson} statistical models for the observations given the auxiliary information.
We note that this is merely a tool to develop our expressions: we run the experiments in Sections~\ref{sec:eval1} and \ref{sec:eval2} with real data.

\subsubsection{Volume leakage.}
Recall that $\Vobs$ is the matrix of observed token volumes, and $\Vaux$ is the matrix of auxiliary keyword volumes.
We use a binomial model to get the coefficients $c$ and $d$.
Namely, assume that, when keyword $\kw{i}$ corresponds to token $\tok{j}$ (denoted $\kw{i}\to\tok{j}$), the number of documents with token $\tok{j}$ (i.e., $\ndocs\cdot\Vobs_{j,j}$) follows a binomial distribution with $\ndocs$ trials and probability given by the auxiliary volume $\Vaux_{i,i}$.
Then, the log-likelihood cost of assigning token $\tok{j}$ to keyword $\kw{i}$ is
\begin{equation}
  d_{i,j} = -\log\Pr\left(\text{Bino}(\ndocs, \Vaux_{i,i}\right) = \ndocs\cdot\Vobs_{j,j})\,,
\end{equation}
where Bino denotes a binomial distribution.
We added a minus sign since we aim at maximizing the log-likelihood, but the linear problem in Alg.~\ref{alg:LAP} is a minimization.
Expanding this probability and ignoring the summands that do not affect the optimization problem, we get
\begin{equation} \label{eq:coefvol}
	d_{i,j} =-\ndocs\left[\Vobs_{j,j}\log(\Vaux_{i,i}) - (1-\Vobs_{j,j})\log(1-\Vaux_{i,i})\right]\,.
\end{equation}
We follow the same approach for the quadratic terms $c_{i,i',j,j'}$, i.e., we assume that when $\kw{i}\to\tok{j}$ and $\kw{i'}\to\tok{j'}$ the number of documents that have both tokens $\tok{j}$ and $\tok{j'}$ follows a Binomial distribution parametrized by $\Vaux_{i,i'}$, and thus the quadratic term is:
\begin{equation} \label{eq:coefvol2}
	c_{i,i',j,j'} =-\ndocs\left[\Vobs_{j,j'}\log(\Vaux_{i,i'}) - (1-\Vobs_{j,j'})\log(1-\Vaux_{i,i'})\right]\,.
\end{equation}

\subsubsection{Frequency leakage with independent queries.}
Recall that $\fobs$ is the vector of observed token frequencies, $\nq$ is the total number of queries issued by the client, and $\faux$ is the vector of auxiliary keyword frequencies.
We use a Poisson model to derive the attack coefficients.
This means that we assume that, when $\kw{i}\to\tok{j}$, the number of times the client sent token $\tok{j}$ follows a Poisson distribution with rate $\nq\cdot\faux_{i}$.
Thus, the cost of assigning $\kw{i}\to\tok{j}$ is $d_{i,j}=-\log\Pr\left(\text{Pois}(\nq\faux_{i})=\nq\fobs_{j})\right)$.
Expanding this expression and ignoring the summands that do not depend on both $i$ and $j$, we get
\begin{equation} \label{eq:coeffreq}
 d_{i,j}=-\nq\cdot\fobs_{j}\cdot\log\faux_{i}\,.
\end{equation}
In this case, the quadratic coefficients are $c_{i,i',j,j'} = 0$ since all queries are independent.

\subsubsection{Frequency leakage with dependent queries.}
\label{sec:Freq}
We extend this approach to the case where we have dependent queries.
Recall that $\rho(\tok{j})$ is the number of times the client queried using token $\tok{j}$, and that $\Fobs_{j,j'}$ contains the number of times the client queried for token $\tok{j'}$ followed by $\tok{j}$, normalized by $\nq(\tok{j'})$.
$\Faux_{i,i'}$ is the probability that the client queried for $\kw{i'}$ followed by $\kw{i}$, as computed from the auxiliary information.
For each free token $\tok{j}\in\tokufree$, we first compute the overall probability that the client sent that token after sending \emph{any other} free token $\tok{j'}\in\tokufree$ ($j\neq j'$):
\begin{equation*}
 \Fobs_{j,\circ}\doteq\frac{\displaystyle\sum_{\tok{j'}\in\tokufree\setminus\tok{j}} \nq(\tok{j'})\cdot\Fobs_{j,j'}}{\displaystyle\sum_{\tok{j''}\in\tokufree} \sum_{\tok{j'}\in\tokufree\setminus\tok{j''}} \nq(\tok{j'})\cdot\Fobs_{j'',j'}}\,.
\end{equation*}
The numerator counts the number of transitions from any other free token to $\tok{j}$ and the denominator is a normalization term so that $\sum_{\tok{j}\in\tokufree}\Fobs_{j,\circ}=1$.
We also define $\nq(\tokufree)$ as the number of times the client queried for a free token other than $\tok{j}$, i.e., $\nq(\tokufree)=\sum_{{\tok{j'}\in\tokufree\setminus\tok{j}}} \nq(\tok{j'})$

Once again we follow a Poisson model to derive the coefficients.
When $\kw{i}\to\tok{j}$, the number of times the client queried for $\tok{j}$ consecutively (i.e., $\rho(\tok{j})\cdot\Fobs_{j,j}$) follows a Poisson distribution with rate $\rho(\tok{j})\cdot\Faux_{i,i}$.
Likewise, the number of times the client queried for another free token followed by $\tok{j}$ ($\rho(\tokufree)\cdot\Fobs_{j,\circ}$) follows a Poisson distribution with rate $\rho(\tok{\circ})\cdot\Faux_{i,\circ}$.
Thus, similar to \eqref{eq:coeffreq}, we set
\begin{equation}  \label{eq:coefFreq2}
	d_{i,j}=-\left[\nq(\tok{j})\cdot\Fobs_{j,j}\cdot\log(\Faux_{i,i})+\nq(\tokufree)\cdot\Fobs_{j,\circ}\cdot\log(\Faux_{i,\circ})\right]\,.
\end{equation}

To get the quadratic terms, we assume that when $\kw{i}\to\tok{j}$ and $\kw{i'}\to\tok{j'}$, the number of times the client sent token $\tok{j}$ after sending $\tok{j'}$ follows a Poisson distribution with rate $\rho(\tok{j})\cdot\Faux_{i,i'}$ (and vice-versa when the client queried for $\tok{j'}$ after $\tok{j}$), which yields
\begin{equation} \label{eq:coefFreq1}
	c_{i,i',j,j'} = - \left[\nq(\tok{j'})\cdot\Fobs_{j,j'}\cdot\log(\Faux_{i,i'})+\nq(\tok{j})\cdot\Fobs_{j',j}\cdot\log(\Faux_{i',i})\right]\,.
\end{equation}

\subsubsection{Leakage combinations}
When the scheme allows both volume and frequency leakage, and the adversary has the corresponding auxiliary information, we combine the coefficients \emph{additively}.
For example, when the scheme leaks the volume and the (independent) query frequencies, we set $d_{i,j}$ as the sum of \eqref{eq:coefvol} and \eqref{eq:coeffreq}, and $c_{i,i',j,j'}$ is simply \eqref{eq:coefvol2}.
The intuition behind this is that, since we are using log-likelihoods, adding coefficients is equivalent to multiplying probabilities, and thus this approach is the maximum likelihood estimator when the volume and frequency leakages are independent.

\subsection{$\name$ vs.~other attacks}
%$\name$ is a query recovery attack that aims at solving a QAP by running iterations in which it solves a LAP ($\LAP$) on randomly chosen keyword and token sets ($\kwufree$ and $\tokufree$).
$\name$ shares certain concepts with previous attacks; we clarify the differences between our attack and related work following.
$\sap$~\cite{oya2021hiding} estimates the keywords of each query by solving a single LAP using a maximum likelihood-based approach.
Therefore, $\sap$ can be written as a single execution of $\LAP$ with only linear ($d$) coefficients.
$\name$ can be seen as an extension of this idea that accommodates quadratic coefficients ($c$) and incorporates the linear program in an iteration loop (Alg.~\ref{alg:attack}).
This results in a more advanced and altogether different attack.
The connection between $\name$ and attacks like $\graphm$~\cite{pouliot2016shadow} and $\ikk$~\cite{islam2012access} is that all these attacks aim at solving a QAP using different heuristics.
$\ikk$ aims at minimizing a Frobenius norm $||\Vaux-\P\Vobs\P^T||_F$, while $\graphm$ adds an additional linear term to this objective function.
$\name$ minimizes a totally different function, characterized by the coefficients $c$ and $d$ we explained above.
$\name$'s function is the first one that can incorporate query correlations into the optimization problem (by setting $d$ as in \eqref{eq:coefFreq1} and \eqref{eq:coefFreq2}).
Aside from the objective functions, the solvers for these three attacks are also completely different: $\graphm$ uses a convex-concave relaxation method, $\ikk$ uses simulated annealing, and $\name$ uses a novel iterative algorithm based on solving linear problems.

\section{Evaluation: Volume-Leaking SSE Schemes}
\label{sec:eval1}

In this section, we compare the performance of our attack against other attacks on SSE schemes that leak the access pattern (S1 and S2 leakage scenarios in Section~\ref{sec:obs}).
We implement our experiments\footnote{\url{https://github.com/simon-oya/USENIX22-ihop-code}} using Python3.8, and run our code in a machine running Ubuntu 16.04, with 1 TB RAM and an Intel(R) Xeon(R) CPU E7 (2.40GHz, 160 CPUs).
We use Oya and Kerschbaum's implementation of $\sap$~\cite{oya2021hiding} and the PATH algorithm~\cite{zaslavskiy2008path} from GraphM package\footnote{\url{https://projects.cbio.mines-paristech.fr/graphm/}} to implement Pouliot and Wright's $\graphm$ attack~\cite{pouliot2016shadow}.

We use five public datasets: Enron, Lucene, Movies, News, and NYTimes.
In Appendix~\ref{app:dataset}, we explain how we obtain and process these datasets, and show their statistical differences (Fig.~\ref{fig:datasets}).
During this processing, we keep the $3\,000$ most popular keywords of each dataset, and remove the rest.
We download the query frequencies of each of those keywords for each week of 2020 from Google Trends \footnote{\url{https://trends.google.com/trends}} using the \texttt{gtab} package~\cite{west2020calibration}.
We found that our results are qualitatively similar across datasets.
Thus, in most of our experiments we show only results in a few datasets, and provide the complete results in Appendix~\ref{app:exp}.

%Figure.~\ref{fig:datasets} shows the keyword volume (percentage of documents in which the keyword appears) of each dataset. \TODO{Maybe show the distribution of keyword volumes instead? Comments that this shows that datasets are different.}

We compare $\name$ with other \emph{statistical-based} attacks, and disregard attacks that rely on ground-truth data~\cite{cash2015leakage, blackstone2020revisiting, damie2021highly, ning2021leap}, since we do not consider this type of leakage in this paper. %\footnote{Even though a more thorough comparison deserves special attention, we achieve higher accuracy than the most recent ground-truth attack when it knows $40\%$ or less of the client's dataset~\cite[Fig.~1]{blackstone2020revisiting}.}
\begin{itemize}
	\item $\sap$~\cite{oya2021hiding} is an attack by Oya and Kerschbaum that uses search and access-pattern leakage and a maximum likelihood approach to solve a LAP \eqref{eq:lap}.
	We use the original implementation by the authors. 
	We set the hypeparameter $\alpha=0.5$ as recommended by the authors.
	\item $\graphm$~\cite{pouliot2016shadow} is the graph matching attack by Pouliot and Wright, which aims at minimizing the function $(1-\alpha)||\Vaux-\P\Vobs\P^T||^2_F- \alpha\Tr(\P^T\mathbf{C})$, where $||\cdot||_F$ denotes the Frobenius norm and $\mathbf{C}$ is a maximum likelihood-based term.
	Following the original work~\cite{pouliot2016shadow}, we use the PATH algorithm~\cite{zaslavskiy2008path} to find a solution to this problem.
	We use $\alpha=0.5$, since we empirically found it performs best.
	%\item $\lgraphm$ is a variation of $\graphm$ that we also try, where we use the logarithm of the term C 's maximum likelihood term is orders of magnitude smaller than the Frobenius norm, and therefore it does not help in the optimization problem.
	%$\lgraphm$ is a modification of $\graphm$ where we use the \emph{logarithm} of this maximum likelihood term (i.e., $\log(\mathbf{C})$) which intuitively will give more weight to this term.
	\item $\ikk$~\cite{islam2012access} aims at minimizing $||\Vaux-\P\Vobs\P^T||_F$ using simulated annealing~\cite{kirkpatrick1983optimization}.
	We set the initial temperature parameter of simulated annealing to $T=100$, a cooling factor of $p_{cool}=0.99995$, and end the algorithm when $T<10^{-10}$.
	This yields running times close to $\graphm$'s.
	%This algorithm starts from a random assignment $\P$ and an initial temperate $T$.
	%In each iteration, the algorithm proposes a candidate assignment $\P'$ by flipping two matchings in $\P$, and moves into it if it yields a smaller Frobenius norm.
	%It also has a certain probability of moving into suboptimal candidates.
	%In each iteration, the temperature gets multiplied by a cooling factor $p_{cool}<1$.
	%As the temperature decreases, the algorithm rejects suboptimal candidates more often.
	%We use an initial temperature $T=200$ and end the algorithm when $T<10^{-10}$.
	%For fairness, we set the cooling factor $p_{cool}=0.99995$; this value brings $\ikk$'s running time close to $\graphm$'s.
	\item $\freq$ is the frequency attack by Liu et al.~\cite{liu2014search}, which uses exclusively frequency information, and simply maps each query token to the keyword whose frequency is closest in Euclidean distance.
\end{itemize}
We tested other generic algorithms for the QAP applied to $||\Vaux-\P\Vobs\P^T||_F$, namely the spectral algorithm by Umeyama~\cite{umeyama1988eigendecomposition} ($\umeyama$) and the fast projected fixed-point algorithm by Lu et al.~\cite{lu2016fast} ($\fgraphm$).
We found these algorithms provide significantly lower accuracy than the other attacks in this list, so we omit them from the paper.
	
In our experiments, we vary the keyword universe size ($\nkw$) and dataset size ($\ndocs$).
Given a value of $\nkw$, we simply build the keyword universe in each run of the experiment by selecting $\nkw$ keywords at random from the set of $3\,000$.
Given a value of $\ndocs$, we sample $\ndocs$ documents from the dataset at random to build the client's dataset.
We also give the adversary $\ndocsaux$ \emph{non-indexed} documents selected at random (i.e., they are not in the client's dataset) as auxiliary information to compute $\Vaux$.
We repeat all of our experiments 30 times, measuring the query recovery accuracy and the running time of the attacks.
We run each attack in a single thread, ensuring that our running time comparisons are fair.
Shaded areas in our plots represent the $95\%$ confidence intervals for the average accuracy.

\subsection{Volume-only leakage attacks}

We consider the setting where the adversary has auxiliary volume information only, and relies on access pattern leakage to estimate the keywords of each query.
Since we do not consider frequency information, we set the coefficients of $\name$ as in \eqref{eq:coefvol} and \eqref{eq:coefvol2}.
We evaluate attacks that use volume ($\name$, $\sap$, $\ikk$, $\graphm$) and exclude $\freq$ from these experiments.

We first consider the case where the adversary observes the access pattern of each query token during initialization (S1, $\ntok=\nkw$). 
Later, we consider the case where only the access pattern of the queried tokens is leaked (S2, $\ntok\leq\nkw$).

\begin{figure}[t]
\centering
\includegraphics[width=0.75\linewidth]{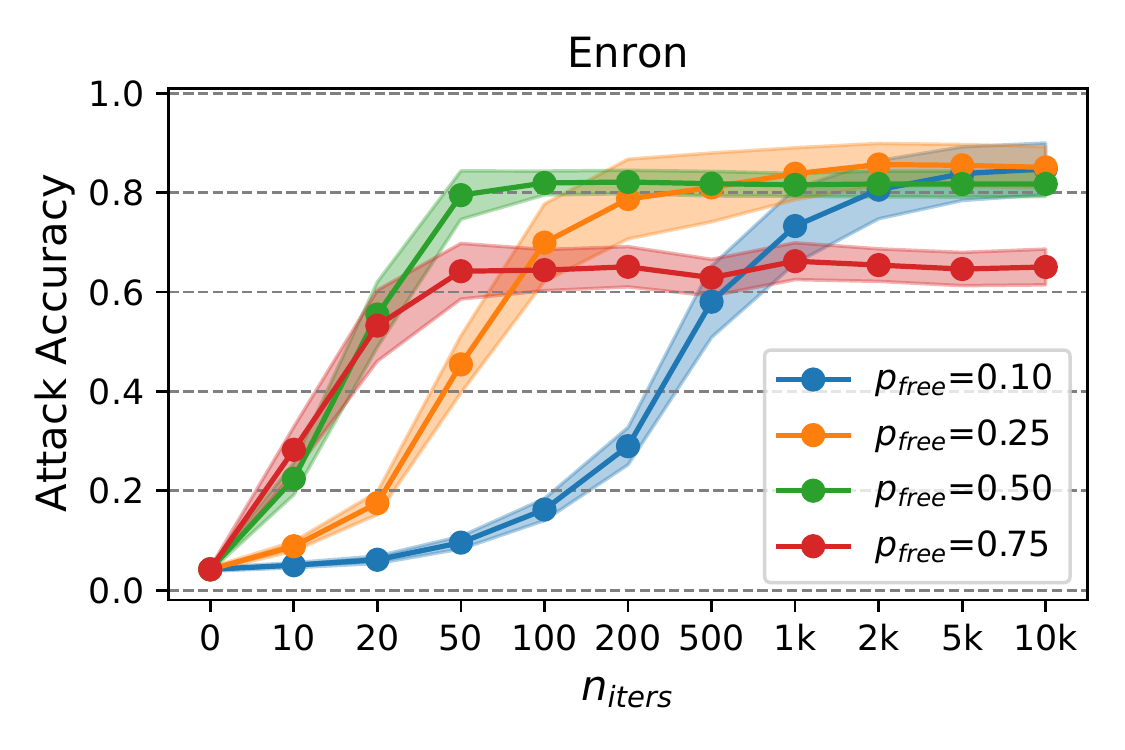}
\caption{Evolution of $\name$'s accuracy with the number of iterations and different $\pfree$ values in Enron dataset. (S1)}
\label{fig:exp0}
\end{figure}

\paragraph{$\name$ parametrization (S1).}
We perform an initial experiment to understand how the number of iterations $\niters$ and the percentage of free tokens $\pfree$ affect the performance of $\name$.
Figure~\ref{fig:exp0} shows the attack accuracy (percentage of correctly guessed query tokens) with the number of iterations, for different values of $\pfree$, in Enron dataset with $\nkw=500$ keywords, $\ndocs=20\,000$ client documents, and $\ndocsaux=5\,000$ auxiliary information documents for the adversary.
%The results are qualitatively similar in the other datasets.
%The lines show the average accuracy and the shaded areas represent the 95\% confidence intervals for this average.
We see that the accuracy increases with the number of iterations and that it converges in all cases, reaching a point where increasing $\niters$ does not yield further improvement.
With $\pfree=0.25$, the attack grows from an accuracy of $4.1\%$ before iterating to an accuracy of $78.7\%$ at 200 iterations, after running for only 38 seconds.
Larger $\pfree$ values allow the algorithm to converge faster, since the algorithm re-computes more assignments per iteration in those cases.
However, a small $\pfree$ increases the number of quadratic terms that are exploited in each iteration, and thus yields higher asymptotic accuracy.
%However, smaller $\pfree$ values yield a higher asymptotic accuracy, since a small $\pfree$ implies that a larger number of quadratic terms are exploited in each iteration.
We also note that smaller $\pfree$ implies faster running times, since each iteration of $\name$ uses the Hungarian algorithm which has a cost that is $O(\nkw\cdot\ntok' + \ntok'^2\cdot\log\ntok')$~\cite{fredman1987fibonacci}, where $\ntok'=\lceil \ntok\cdot\pfree\rceil$.
The running time of $\name$ for $\pfree=0.1, 0.25, 0.5$, and 0.75 in Enron was 0.14, 0.19, 0.34, and 0.46 seconds per iteration, respectively.
We use $\pfree=0.25$ in the remainder of the paper, since it offers a good trade-off between convergence speed and asymptotic performance.

%\subsubsection{Attack comparison in S1}

\begin{figure*}[t]
\centering
\includegraphics[width=0.95\linewidth]{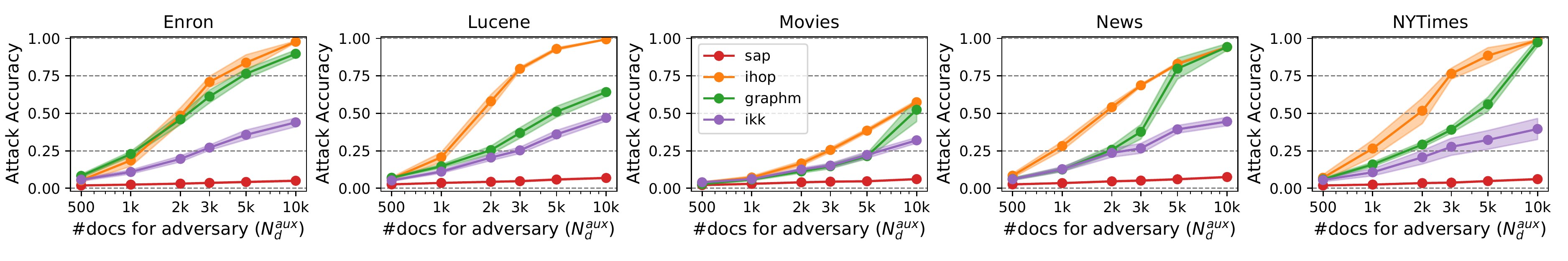}\\
\caption{Attack comparison in all datasets vs.~the number of non-indexed documents of the adversary ($\ndocsaux$) (S1)}
\label{fig:exp10}
\end{figure*}

\paragraph{Attack comparison with $\ndocsaux$ (S1).}
We compare the performance of the different attacks in the S1 setting.
We set the number of keywords to $\nkw=500$, since $\graphm$ is computationally expensive.
We build the client's dataset with $\ndocs=20\,000$ documents, and vary the amount of documents we give the adversary as auxiliary information $\ndocsaux\in[500,10\,000]$.
Figure~\ref{fig:exp10} shows the results (we used $\name$ with $1\,000$ iterations and $\pfree=0.25$).
We see that the attacks' accuracy increases as the quality of the auxiliary information grows.
The accuracy depends on the dataset, but we see that $\name$ achieves higher accuracy than other attacks except in the particular case of Enron with $\ndocsaux=500$ and $1\,000$.
The average running times of the attacks in Enron, which are similar across $\ndocsaux$, are 33 seconds for $\sap$, 231 seconds for $\name$, 3.3 hours for $\graphm$, and 2.3 hours for $\ikk$.

%Next, we set the auxiliary information size to $\ndocsaux=10\,000$ and increase the client's dataset size $\ndocs=30k, 50k, 100k, 200k$.
%We run this experiment in NYTimes dataset only, since it's the only one that has enough documents, and show the results in Figure~\ref{fig:exp12}.
%\TODO{Get this.}

\begin{figure}[t]
\centering
\includegraphics[width=0.95\linewidth]{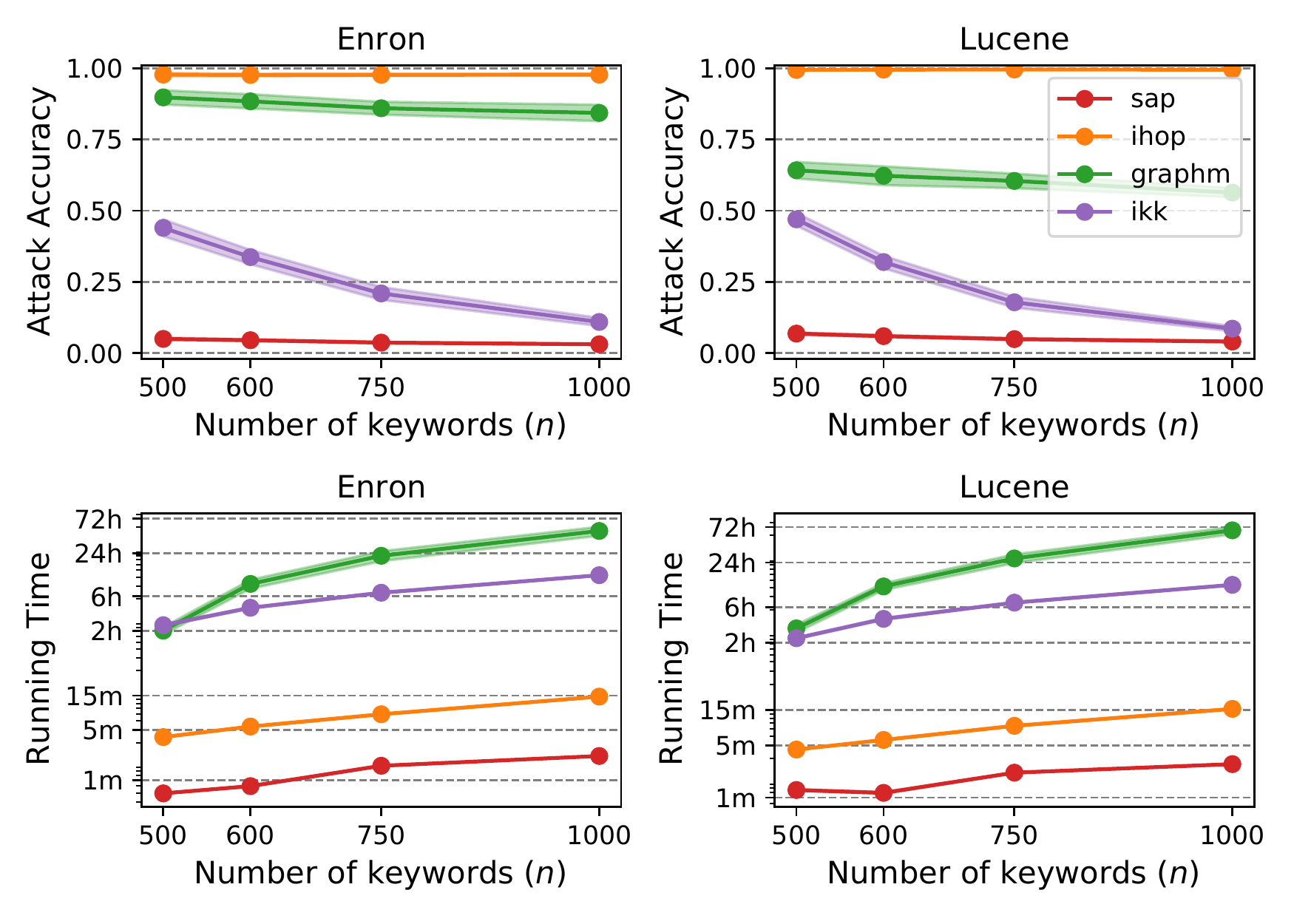}\\
\caption{Attack comparison vs.~number of keywords ($n$) (S1)}
\label{fig:exp11}
\end{figure}

\paragraph{Attack comparison with $\nkw$ (S1).}
Next, we study how the keyword universe size affects the performance of the attacks.
We set $\ndocs=20\,000$ and $\ndocsaux=10\,000$, and progressively increase $\nkw$ from $500$ to $1\,000$.
Figure~\ref{fig:exp11} shows the results in Enron and Lucene datasets. %\footnote{Our experiments are qualitatively similar across all five datasets.
%We provide the results of all of our experiments for all datasets in the full version of the paper.}
The top plots show that the attack accuracy remains steady for $\name$, slightly decreases for $\graphm$, and significantly decreases for $\ikk$.
The bottom plots show the running times in logarithmic scale.
We see that $\graphm$ and $\ikk$ quickly become unfeasible as the keyword universe size increases: with $\nkw=1\,000$ keywords, $\graphm$ takes around 3 days to finish, while $\name$ ends in 15 minutes and achieves higher accuracy.

\begin{figure}[t]
\centering
\includegraphics[width=0.95\linewidth]{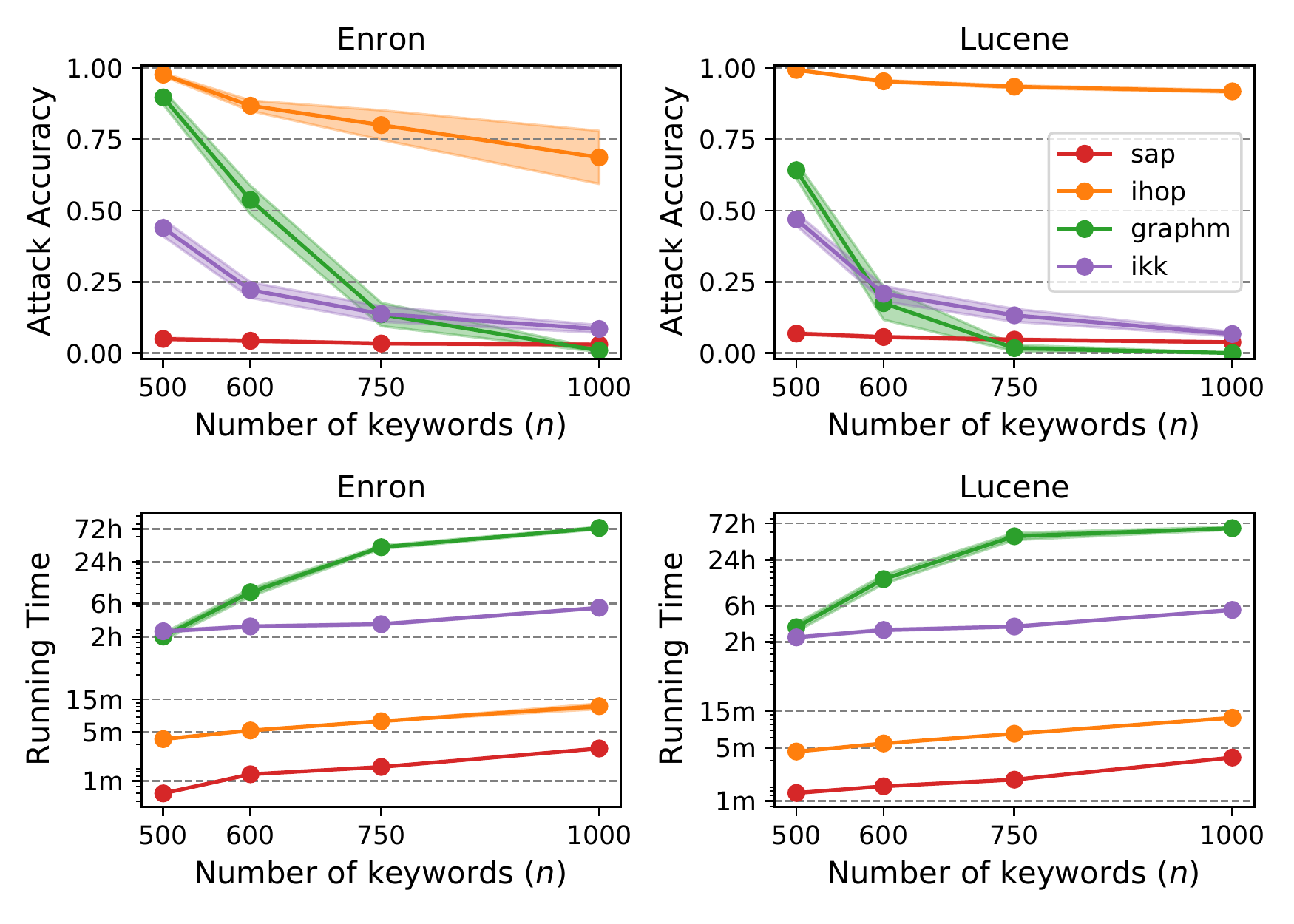}\\
\caption{Attack comparison vs.~number of keywords ($n$) when the adversary sees $\ntok=500$ distinct tokens (S2)}
\label{fig:exp11_fixed}
\end{figure}

\paragraph{Attack comparison with $\nkw$ and fixed number of observed tokens $\ntok$ (S2).}
We repeat the previous experiment in the setting where only the access patterns of the queried keywords are leaked (S2).
We fix the number of distinct keywords queries (i.e., distinct tokens observed) to $\ntok=500$, and increase $\nkw$ as above.
This means that the adversary observes the access pattern of $\ntok=500$ query tokens and has to guess their corresponding keywords from a larger set $n\geq 500$.
We show the results in Figure~\ref{fig:exp11_fixed}.
As expected, the accuracy of all attacks decreases in this case compared to when all access patterns are observed.
$\graphm$ is particularly affected by this.
This is due to a limitation of the objective function that $\graphm$ minimizes ($||\Vaux-\P\Vobs\P^T||^2_F$ unfairly penalizes unassigned keywords when $n\geq m$).
$\name$ achieves the highest accuracy among all attacks in all datasets, and a running time orders of magnitude below $\ikk$ and $\graphm$.

\subsection{Volume and frequency leakage attacks}
\label{sec:freq}

For the next set of experiments, we consider both volume and frequency leakage when the client sends queries independently.
For each experiment, after selecting the keyword universe $\kwu$ as above, we take the frequency data from Google Trends for the keywords in $\kwu$ over the span of 2020.
Then, we use the average frequencies of the first half of this year as the auxiliary information $\faux$ and generate keywords independently at random following the average query frequencies of the second half of the year (this affects the observed frequencies of tokens $\fobs$).
Since we consider independent query generation, the adversary uses $\fobs$ and $\faux$ in their attack.
We set the coefficients of $\name$ as the summation of \eqref{eq:coefvol} and \eqref{eq:coeffreq}, and use $\niters=1\,000$ and $\pfree=0.25$.
Besides the attacks we considered above, we also evaluate $\freq$.

%We vary the total number of queries $\nq$.
%We also consider $\freq$, $\sap$ with $\alpha=0.5$.
%We do not consider $\ikk$ or $\graphm$, since they cannot exploit frequency information and we have seen $\name$ outperforms them even without this information.

\paragraph{Attack comparison with $\nq$ (S2).}

\begin{figure}[t]
\centering
\includegraphics[width=0.95\linewidth]{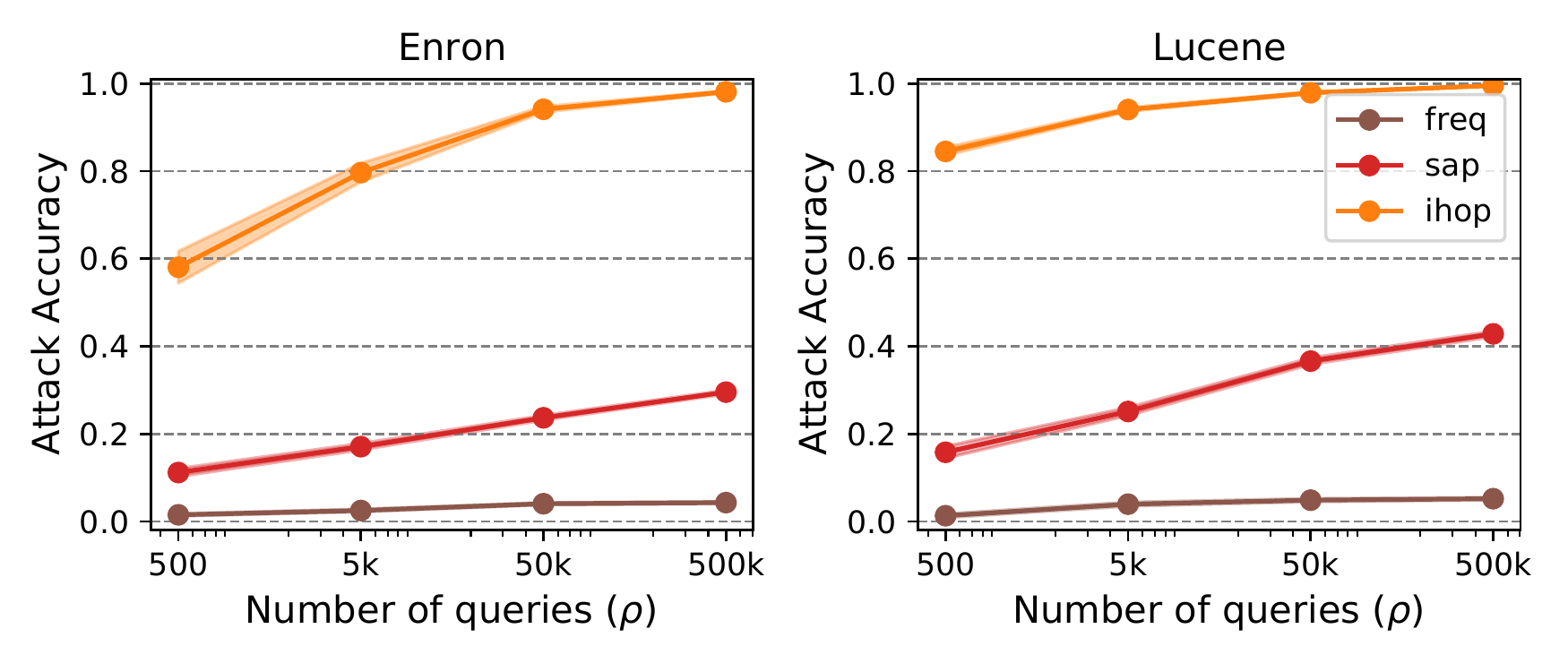}\\
\caption{Comparison of attacks that use frequency leakage, with $\nkw=3\,000$ (S2)}
\label{fig:s2}
\end{figure}

We first compare the attacks that rely on frequency information ($\name$, $\sap$, and $\freq$).
We set $\nkw=3\,000$, since these attacks are efficient, and vary the number of queries $\nq$.
Figure~\ref{fig:s2} shows the attack accuracy (recall that this is the percentage of \emph{distinct} query tokens whose underlying keyword is correctly guessed by the attack).
$\name$ comfortably beats both $\sap$ and $\freq$ ($+20\%$ more accuracy than these attacks in all cases), which are the state-of-the-art in this setting.
This is because $\name$ is the only attack that can exploit both frequency and volume co-occurrence information at the same time.
%\TODO{Give some time information?}
%Our attack's running time sits around 100 seconds, slower than both $\freq$ and $\sap$, but around two orders of magnitude faster than $\graphm$.

\paragraph{Access-pattern obfuscation defenses (S2).}
We evaluate all attacks against two access-pattern obfuscation defenses: $\clrz$~\cite{chen2018differentially} and $\osse$~\cite{shang2021obfuscated}.
$\clrz$ randomly adds and removes keywords from documents before outsourcing the database.
The False Positive Rate (FPR) of $\clrz$ is the probability that it adds a keyword to a document that does not have it, and the True Positive Rate (TPR) is the probability that it keeps a keyword in a document that has it.
$\clrz$ obfuscation occurs before outsourcing the database, and therefore querying for the same keyword twice produces the same (obfuscated) access pattern.
$\osse$ achieves the same effect during query time: querying for a particular keyword can return documents that do not contain that keyword and miss some documents that do contain the keyword. 
Querying for the same keyword twice yields (with high probability) different access patterns in $\osse$, providing a certain level of search pattern privacy.

\begin{figure}[t]
\centering
\includegraphics[width=0.95\linewidth]{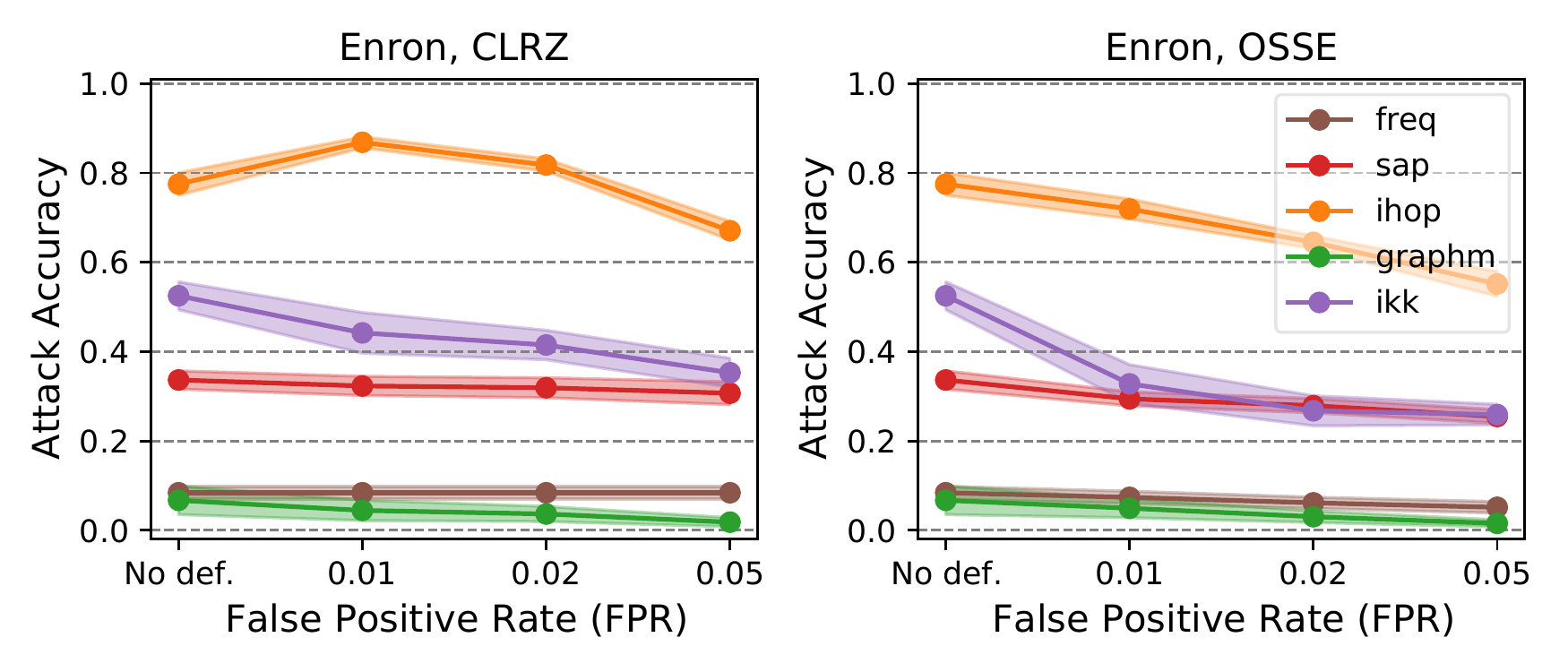}\\
\caption{Attack comparison vs.~$\clrz$ and $\osse$ in Enron dataset ($\nkw=500$, $\nq=500$) (S2)}
\label{fig:s21}
\end{figure}

%Therefore, we focus on $\clrz$ and $\osse$.
We assume the adversary knows the defense parameters (TPR and FPR) and explain in Appendix~\ref{app:adapt_osse} how we adapt all the attacks against these defenses.
We set $\nkw=500$, since $\graphm$ is too slow for larger keyword universes, and set the number of queries to $\nq=500$.
We set TPR=0.9999 and vary FPR for both defenses.
Figure~\ref{fig:s21} shows the accuracy of the attacks against $\clrz$ and $\osse$ in Enron dataset.
The figure shows that $\osse$ is a stronger defense than $\clrz$, but we observe a smaller difference between these defenses than Shang et al.~\cite{shang2021obfuscated}, since we have adapted the attacks against $\osse$ more efficiently.
More importantly, we see a large accuracy gap between $\name$ and the other attacks.\footnote{We observe an increase in the accuracy of $\name$ when FPR=0.01.
We only observed this effect in Enron and Movie datasets, and we conjecture it is an effect of these datasets' distribution and how $\name$'s objective function interacts with it.}

%Regarding attacks, we consider $\sap$, since it has been previously adapted against $\clrz$, as well as our attack $\name$.
%We assume the adversary knows the defense parameters (TPR and FPR) and explain in Appendix~\ref{app:adapt_osse} how we adapt $\sap$ and $\name$ against these defenses.

%This means that we assume the adversary knows the TPR and FPR of the defense, and modifies the constants $c$ and $d$ using the maximum likelihood-based approach accordingly.
%Shang et al.~\cite{shang2021obfuscated} explain how to adapt different attacks against $\clrz$ and $\osse$, and Oya and Kerschbaum~\cite{oya2021hiding} already show that $\sap$ is effective against $\clrz$.
%In the Appendix~\ref{app:adapt}, we explain how we adapt $\name$ and $\sap$ against both $\clrz$ and $\osse$.

\paragraph{Other defenses: discussion.}
We only considered the $\clrz$ and $\osse$ defenses because they affect the observed volumes $\Vobs$ without modifying the general leakage format we assume in this paper.
We did not consider other defenses like the proposal by Patel et al.~\cite{patel2019mitigating} or SEAL~\cite{demertzis2020seal} because they pad the documents that are returned to each keyword independently, but do not consider how to pad volume co-occurrence.
This works when each document has a single keyword (or attribute), but not in the \emph{keyword query} scenario we consider in this paper unless the client replicates each document once per each of its keywords, which requires an unfeasible storage cost. 
We also disregarded $\pancake$~\cite{grubbs2020pancake} in this section since it hides frequencies but not volume (we consider this defense in the following section), and SWiSSSE~\cite{gui2020swissse} since it is an altogether more complex SSE scheme that does not fit our description in this paper, and we believe deserves individual attention.
We note that a purely ORAM-based defense can protect against $\name$, but these defenses come at a high overhead cost; our attack is effective against efficient defenses like $\clrz$~\cite{chen2018differentially}.

\section{$\name$ against $\pancake$ with Query Dependencies}
\label{sec:eval2}

In this section, we evaluate the performance of $\name$ in a setting where the client's queries are correlated, i.e., querying for a particular keyword $\kw{i}$ affects the probability of the keyword chosen for the next query.
This query model is interesting because humans rarely make independent decisions, and this type of query correlations have not been considered in attacks against SSE schemes before (except briefly by Grubbs et al.~in their appendix~\cite{cryptoeprint:2020:1501}).
As we explained in Sec.~\ref{sec:Freq}, $\name$ can take these correlations into account.
We consider a case where there is frequency-only leakage (S3) so that the attack's success relies exclusively on exploiting these query dependencies.
A particular case of frequency-only leakage is when the client simply queries for individual documents using their identifier or a ``document title''.

%In this section, we configure $\name$'s coefficients as in \eqref{eq:coefFreq1} and \eqref{eq:coefFreq2}.

Besides showing the effectiveness of $\name$ with query correlations, the main contribution of this section is an evaluation of $\pancake$, the frequency hiding defense by Grubbs et al.~\cite{grubbs2020pancake}, in the presence of query dependencies.
We provide an overview of $\pancake$ below, then explain how we adapt $\name$ against $\pancake$, and finally introduce our experimental setup and show our results.

\subsection{Overview of $\pancake$}
\label{sec:pancake}

$\pancake$~\cite{grubbs2020pancake} is a system that hides the frequency access of key-value datasets by using a technique called frequency smoothing.
$\pancake$ uses \emph{document replication} and \emph{dummy queries} to ensure that the access frequency to the key-values in the dataset is uniform.
For compatibility with our notation, we consider the case where the keyword universe and dataset have the same size ($\nkw=\ndocs$), and without loss of generality keyword $\kw{i}$ matches only document $d_i$ ($i\in[\nkw]$).

\begin{figure}
\begin{center}
\def\svgwidth{\linewidth} 
{
%\fontsize{7.5pt}{9.5pt}
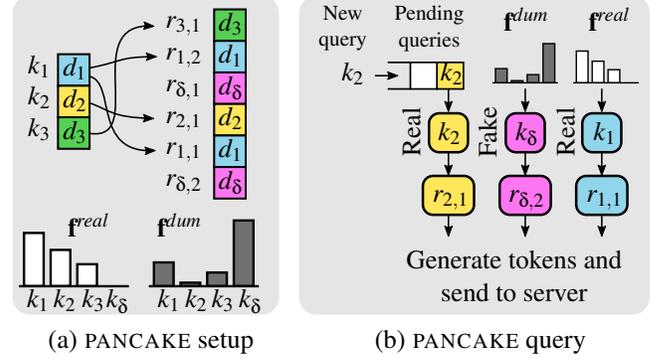
}
\caption{Overview of $\pancake$'s setup and queries.\label{fig:pancake}}
\end{center}
\end{figure}

We use Figure~\ref{fig:pancake} to summarize how $\pancake$ works.
%Here we provide a simplified overview of the concepts of $\pancake$ that are needed for our paper, and refer to the original paper for full details~\cite{grubbs2020pancake}.
In the figure, $\nkw=\ndocs=3$.
Let $\freal$ be a vector of length $\nkw+1$, where its $i$th entry $\freal_i$ contains the real query frequency of keyword $\kw{i}$, and $\freal_{\nkw+1}=0$ represents the real query frequency of a dummy keyword-document pair $\kw{\delta}$-$d_\delta$.
We assume that the client knows $\freal$ for simplicity (this only benefits the client and not the attack).
The client creates $R(i)=\lceil \freal_i / \nkw \rceil \geq 1$ replicas for each document $d_i$ ($i\in[\nkw]$), and additionally creates $R(\nkw+1)=2\nkw-\sum_{i=1}^\nkw R(i)$ dummy replicas $d_\delta$.
This makes a total of $\nrep$ document replicas, that are encrypted and sent to the server (in a random order).
In Figure~\ref{fig:pancake}a, the client creates two replicas for $d_1$, one for $d_2$ and $d_3$, and two dummy replicas $d_\delta$.
We refer to the $\ell$th replica of $\kw{i}$ by $\rep{i,\ell}$, with $\ell\in[R(i)]$.
The client computes a vector of dummy frequencies $\fdum_i=R(i)/\nkw-\freal_i\geq 0$ for each $i\in[\nkw+1]$.
For simplicity, we assume that the client knows the mapping of keywords to replicas and stores $\freal$ and $\fdum$ locally, and refer to the original paper for more advanced details~\cite{grubbs2020pancake}.

When the client wishes to query for a keyword $\kw{i}$, she places $\kw{i}$ inside a buffer of pending queries, and flips \emph{three} unbiased coins (Fig.~\ref{fig:pancake}b).
For each coin flip: if it shows heads, the client queries for the next keyword in her buffer of pending queries (e.g., $\kw{2}$ in~\ref{fig:pancake}b).
If the buffer is empty, she simply samples a keyword from $\freal$ and queries for it (e.g., $\kw{1}$ in Fig.~\ref{fig:pancake}b).
If the coin shows tails, the client samples a keyword from $\fdum$ and queries for it (e.g., $\kw{\delta}$ in Fig.~\ref{fig:pancake}b).
To query for a keyword, the client selects one of its replicas at random and generates a token that retrieves the document associated with that replica.

The probability that the client sends a query for keyword $\kw{i}$ is therefore $\frac{1}{2}\freal_i + \frac{1}{2}\freal_i=R(i)/2\nkw$.
Since the client chooses one of the $R(i)$ replicas at random when querying for $\kw{i}$, the access frequency of any replica is $1/2\nkw$.
Since each query token corresponds to one replica, the frequency of each token $\tok{j}$ ($j\in[2\nkw]$) is also $1/2\nkw$ (uniform); i.e., query tokens are indistinguishable based on their frequencies.

%\begin{figure*}[t]
%\centering
%\includegraphics[width=0.95\linewidth]{pancake.png}\\
%\caption{\TODO{This is just a template!}.}
%\label{fig:pancake}
%\end{figure*}

\subsection{Query dependencies against $\pancake$}

$\pancake$ ensures that each document replica is accessed with probability $1/2\nkw$.
However, when there are dependencies between the keywords that the client chooses, $\pancake$ does not hide these dependencies.
Grubbs et al.~\cite{grubbs2020pancake} provide strong evidence that $\pancake$ obfuscates query correlations.
Here, we perform another study of $\pancake$ against dependent queries, and discuss our findings against the ones by Grubbs et al.~\cite{grubbs2020pancake} in Section~\ref{sec:disc}.
We consider the case where the client follows a Markov model to choose the keywords of her queries, and use $\Freal$ to denote the $\nkw\times\nkw$ Markov matrix that characterizes the client's querying behavior.
We assume that the Markov process is irreducible and aperiodic so that it has a unique stationary distribution, $\freal$.
This vector ($\freal$) contains the probability that the client queries for each keyword (at any point in time) and can be computed analytically from $\Freal$.
The client uses $\freal$ as input to $\pancake$ to compute the replicas and dummy distribution.
Even though the query frequency of each token is $1/2\nkw$, the correlations between real queries cause correlations between query tokens which the adversary can leverage for query recovery (we show an example of this in Fig.~\ref{fig:pancake_app}, in the appendix).

Now we explain how we adapt our attack against $\pancake$.
When the client queries $\nq$ times, she sends a total of $3\nq$ tokens.
The adversary observes the sequence of tokens $\obs=[(\tok{})_1, \dots, (\tok{})_{3\nq}]$ and groups them, consecutively, into triples.
Then, the adversary builds the $\nrep\times\nrep$ matrix of observed frequencies $\Fobs$ by considering every two consecutive triples and counting each transition from a token in the first triple and a token in the second triple. 
This matrix is then normalized by columns, which ensures it is left-stochastic.
%Figure~\ref{fig:triples} shows a toy example of building $\Fobs$.

The adversary has an $\nkw\times\nkw$ auxiliary information matrix $\Faux$, which captures the client's querying behavior (ideally, this matrix is close to $\Freal$).
The adversary computes the stationary distribution of $\Faux$ (i.e., $\faux$), and uses it to get the expected number of replicas for each keyword and the dummy profile that the client is using.
Note that if $\Faux$ is far from $\Freal$, the adversary's belief of the number of replicas per keyword and the dummy profile might be very far from the true ones.
%The adversary computes the matrix of expected frequencies between replicas, given $\Faux$, that we denote by $\Fexp$.
%We explain how to compute $\Fexp$ in Appendix~\ref{app:adapt_pancake}.
In Appendix~\ref{app:adapt_pancake}, we explain how the adversary can compute the matrix of \emph{expected} frequencies between replicas given $\Faux$, which we denote by $\Fexp$.
The adversary uses the matrix of observed token frequencies $\Fobs$ and expected replica frequencies $\Fexp$ to match tokens to replicas.
This in turn yields a matching between tokens and keywords.
This means that we adapt $\name$ against $\pancake$ by setting its $c$ and $d$ coefficients as in \eqref{eq:coefFreq1} and \eqref{eq:coefFreq2}, using $\Fexp$ instead of $\Faux$ in these expressions.

%Let $\P^{\rep{}\to\kw{}}$ be an (arbitrary) mapping between each replica and its corresponding keyword, according to the number of replicas computed from $\faux$.
%The adversary computes the $\nrep\times\nrep$ matrix $\Fexp\doteq\text{E}\{\Fobs|\Faux,\P^{\rep{}\to\kw{}}\}$.
%This matrix represents the same statistical information as the matrix of observed transition counts $\Fobs$, but where $\Fobs$ refers to tokens, $\Fexp$ refers to replicas.
%In Appendix~\ref{app:adapt_pancake}, we give an expression for $\Fexp$ and provide examples of matrices $\Freal$, $\Fobs$, and $\Fexp$, intuitively explaining why query correlations allow query recovery attacks in $\pancake$.
%To adapt $\name$ against $\pancake$, we set its $c$ and $d$ coefficients as \eqref{eq:coefFreq1} and \eqref{eq:coefFreq2}, using $\Fexp$ instead of $\Faux$ in these expressions.
%The result of this attack is a mapping between tokens and replicas, which the adversary transforms to a mapping between tokens and keywords following $\P^{\rep{}\to\kw{}}$.

\subsection{Experimental Results}
\label{sec:eval_pancake}
 
For our evaluation, we consider the case where the client's database is an encyclopedia that she wishes to securely store on a server.
Each document contains information about a particular topic, identified by its keyword.
To build our datasets and compute the transition probabilities between the queries, we use data from the Wikipedia Clickstream dataset.\footnote{\url{https://dumps.wikimedia.org/other/clickstream/}}
We summarize our approach here, and provide full details in Appendix~\ref{app:wiki}.
Using PetScan,\footnote{\url{https://petscan.wmflabs.org}} we get all Wikipedia pages under the five categories  \emph{`privacy'}, \emph{`security'}, \emph{`cryptography'}, \emph{`politics'}, and \emph{`activism'}, including pages in subcategories.
This yields a total of $5\,573$ pages.
Many of these pages have hyperlinks that allow a user to browse from one to another.
We download the number of times Wikipedia users clicked on these hyperlinks to transition between these pages in 2020.
We build five keyword universes of size $\nkw=500$, each focused on one of the five categories above.
For each of these keyword universes, we use the transition counts to build two $500\times 500$ transition matrices: one with data from January 2020 to June 2020 ($\Fold$), and another one with data from July 2020 to December 2020 ($\Fnew$).
%We note that the dataset information varies significantly between months, and the average distance per column (i.e., per transition profile) between $\Fold$ and $\Fnew$ is $\approx 0.40$, which is a large value (the maximum absolute difference between two probability distributions is 2).
We use $\Fnew$ to generate the client's real queries, i.e., $\Freal=\Fnew$.
We evaluate both the case were the adversary has low-quality auxiliary information ($\Faux=\Fold$, denoted $\aux\downarrow$ in the plots) and high-quality information ($\Faux=\Fnew$, denoted $\aux\uparrow$).

We consider two values of the total number of queries: $\rho=100\,000$ and $\rho=500\,000$.
These are large numbers, because the datasets contain $\nkw=500$ keywords, which generates $1\,000$ replicas with $\pancake$, and thus there are $1\,000$ possible query tokens.
These tokens are all queried with the same probability (this is guaranteed by $\pancake$), and thus with $\rho=100\,000$ (resp., $\rho=500\,000$) the server observes $\approx 300$ (resp., $\approx 1\,500$) transitions from (or to) each query token, which we believe is not a large number.
We note that attacks that exploit query dependencies need such large number of queries to succeed (except in cases where query dependencies are unusually high).
%We admit that the total number of queries $\rho$ is larger than in our previous experiments and that the effectiveness of $\name$ using query correlation information is subject to having observed a large number of queries.\smargin{this sentence is perhaps a bit weird? maybe we should have a discussion section later talking about what Grubbs does in their comparison and the difference with ours, and also this}

\begin{figure*}[t]
\centering
\includegraphics[width=0.95\linewidth]{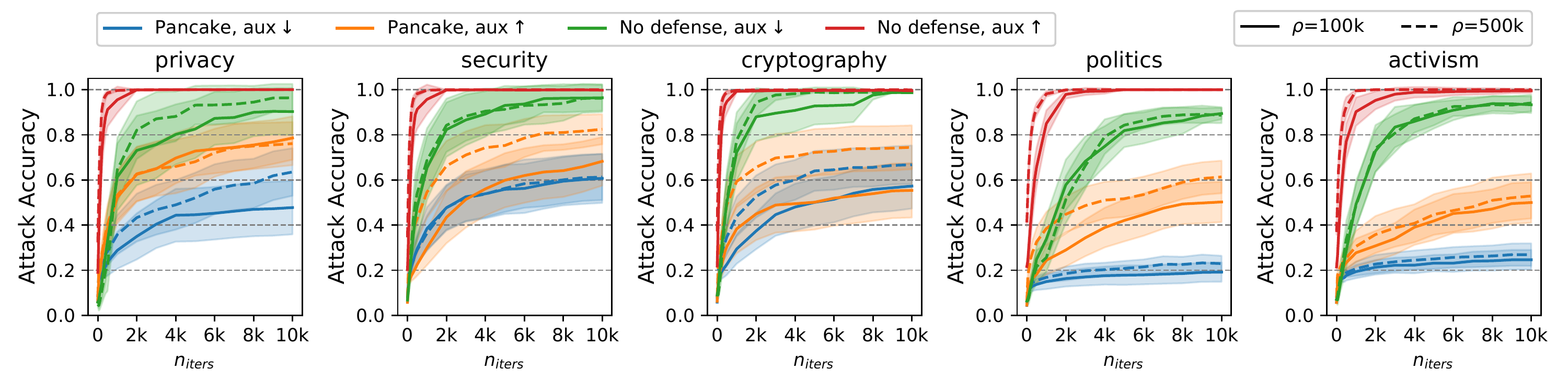}\\
\caption{Accuracy of $\name$ against $\pancake$ with the number of iterations $\niters$, for different category-based datasets and number of queries $\nq$. $\aux\uparrow$ and $\aux\downarrow$ denote high-quality and low-quality auxiliary information, respectively. (S3)}
\label{fig:pancake_cat}
\end{figure*}

We evaluate $\name$ with $\pfree=0.25$ and use $\niters=10\,000$.
Figure~\ref{fig:pancake_cat} shows the accuracy of $\name$ vs.~the number of iterations $\niters$, for each of the five category-based keyword universes.
As before, the accuracy in these plots is the percentage of query tokens (out of the $2\nkw$ tokens) for which the adversary guesses the underlying keyword correctly.
In each plot, the title shows the category, continuous and dashed lines denote $\nq=100\,000$ and $\nq=500\,000$, respectively, and each color denotes which defense is used and the quality of auxiliary information as explained above.
We see that the attack accuracy and its evolution with $\niters$ varies significantly depending on the category (i.e., the underlying Markov model $\Freal$ affects the algorithm's convergence speed and accuracy).
In most cases, $\pancake$ significantly reduces the attack accuracy (blue vs.~green lines, and orange vs.~red lines).
As expected, increasing the number of queries observed usually improves the attack.
The exception to this is the green lines for some categories: this is because for these categories the imperfect auxiliary information ($\aux\downarrow$) is misleading, so observing more queries misleads the attack even further.
We see a slower convergence for $\name$ compared to Fig.~\ref{fig:exp0}; this is because the attack has volume information in Fig.~\ref{fig:exp0}, which makes the linear coefficients $d$ very helpful even when the fixed assignment in an iteration $\Pfixed$ contains many incorrect matchings.
Our attack in this frequency-only setting (Fig.~\ref{fig:pancake_cat}) relies mostly on the quadratic terms $c$, which are only truly helpful when the current assignment $\P$ is already accurate.
This makes the attack's convergence slower.

\begin{figure}[t]
\centering
\includegraphics[width=0.75\linewidth]{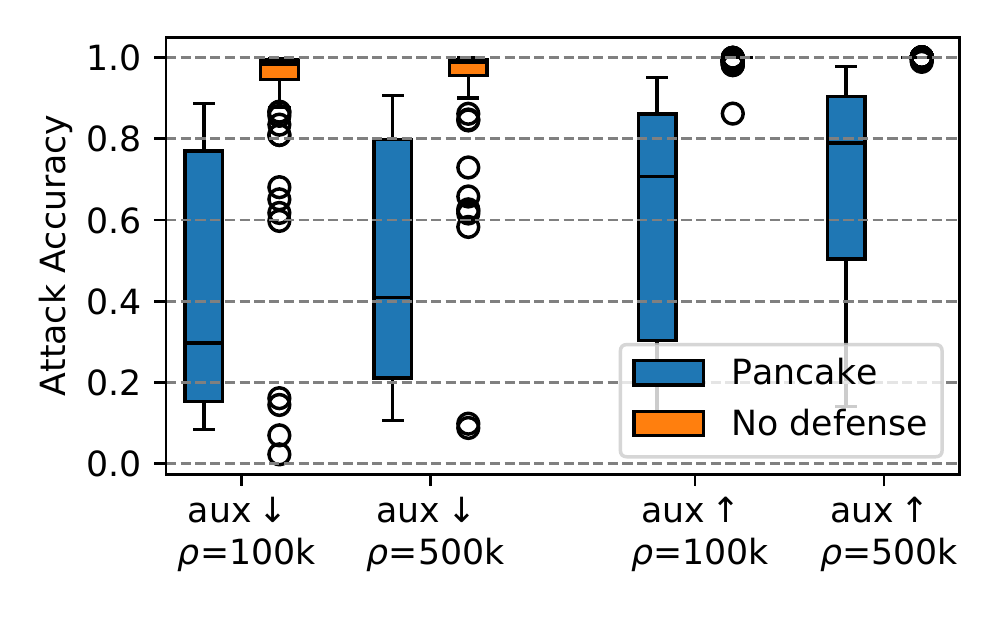}\\
\caption{Summary of results of $\name$ against $\pancake$ with query dependencies. (S3)}
\label{fig:pancake_box}
\end{figure}

We summarize the results of our experiment in Figure~\ref{fig:pancake_box}.
Here, each box represents the accuracy of the attack against \emph{all} categories (30 accuracy values for each category, 5 categories) in the last iteration of the attack ($\niters=10\,000$).
This figure shows more clearly the advantage of using $\pancake$ over no defense when there are query dependencies.
For $\nq=100\,000$ queries, using $\pancake$ decreases the average attack accuracy from 93.6\% (no defense) to 41.9\% ($\pancake$), when the auxiliary information is low-quality.
With high-quality auxiliary information, the accuracy decreases from 99.7\% to 60.5\% thanks to $\pancake$. 
Although the decrease is somewhat significant, our results confirm that $\pancake$ is vulnerable to query correlations, which differs from the findings in the preliminary analysis by Grubbs et al.~\cite{cryptoeprint:2020:1501}.
The accuracy gap between the attack without defense and with $\pancake$ is even smaller when the number of queries increases.

%We also ran $\name$ using the individual frequencies $\faux$ and $\fobs$ (i.e., without exploiting correlations), and the accuracy was only between $1\%$ and $11\%$ when the client does not use any defense and $<0.1\%$ when using $\pancake$ (in this case the attack is not better than randomly assigning tokens to keywords).
%This shows that query frequencies can significantly help increase the accuracy of query recovery attacks.
%\smargin{TODO: fix these numbers. We might just remove this paragraph if we need room}

\subsection{Discussion}
\label{sec:disc}

Our experiments show that $\pancake$ might not provide enough protection when there are query dependencies, which differs from the findings by Grubbs et al.~\cite{grubbs2020pancake}.
The reason for this is that Grubbs et al.~consider a case where the client's dataset contains mappings of keywords to document identifiers, and of document identifiers to documents.
The client queries first for a keyword, followed by a query for a document that contains that keyword.
This creates dependencies between pairs of queries.
Their dataset contains $\approx 532\,000$ entries, which results in over one million replicas.
They see that, even when the client issues 10 million queries, the joint distribution of consecutive accesses is almost flat.
This is reasonable, since with one million replicas, 10 million queries might not be enough information for the adversary.
Even with enough queries, carrying out $\name$ in this case would be computationally prohibitive due to the size of the problem.

%Also, we think that in a realistic case the keywords in the case of Grubbs et al.~would also be correlated, so the actual observations might exhibit higher correlations than in their analysis.
%To summarize: $\pancake$ seems to protect against query correlations for large keyword universe and dataset sizes, but we have provided the first piece of evidence that in a setting with 500 keywords $\pancake$ does not provide sufficient protection against query correlations.
%Developing a statistical-based query recovery attack able to deal with large keyword universe sizes and evaluating it against $\pancake$ with a real (correlated) query data is an interesting future research line, that falls outside of the scope of our work.
To summarize: we have provided the first piece of evidence that there are cases where $\pancake$ does not provide sufficient protection against query correlations. 
$\pancake$'s protection against correlated queries thus depends on the keyword universe and dataset sizes, as well as the strength of the correlations.
Developing statistical-based query recovery attacks able to deal with large keyword universe sizes and studying $\pancake$'s protection for different correlation levelsis an interesting future research line. %, that falls outside of the scope of our work.

\section{Conclusion}
We proposed $\name$, a new attack on SSE schemes that uses statistical (non-ground-truth) auxiliary information to recover the client's queries.
Our attack formulates query recovery as a quadratic optimization problem and uses a novel iteration heuristic that relies on efficient optimal linear solvers to find a suitable solution.
$\name$ is the first query recovery attack that can leverage both quadratic volume and frequency terms.

We evaluated our attack on real datasets against SSE schemes that exhibit typical leakage patterns, showing that it outperforms all other statistical-based query recovery attacks.
In four out of the five datasets we consider, $\name$ achieves almost perfect query recovery accuracy ($\approx 100\%$) in the full access-pattern leakage setting when the adversary receives a number of non-indexed documents equal to half the dataset size, while being more than one order of magnitude faster than previous attacks.
When the adversary sees the access patterns of a subset of all the keywords only, $\name$ widely outperforms all other attacks ($92\%$ accuracy in Lucene; previous best achieves $7\%$), while being significantly faster (12 minutes vs.~over 5 hours).
We verified $\name$ consistently outperforms other attacks when frequency information is available, as well as when efficient access-pattern obfuscation techniques are applied.
 %$80\%$\smargin{TODO: fix these numbers} query recovery accuracy in the full access-pattern leakage setting in Enron dataset ($90\%$ in Lucene), while previous attacks are around $50\%$ and more than one order of magnitude slower.
%Against SSE schemes that only leak the access pattern and frequency of the queried keywords, $\name$ achieves an accuracy that is always $+20\%$ higher than the best-performing existing attack, sometimes reaching $90\%$ accuracy.
%We adapt $\name$ against recent access and search pattern-hiding defenses and show that it outperforms the best existing attack against such defenses.
Finally, we demonstrate that $\name$ can exploit query dependencies by adapting our attack against the $\pancake$ frequency-smoothing defense.
Our results in a small dataset confirm that $\pancake$ might not provide sufficient protection against query recoveries and urge for a more thorough analysis of $\pancake$ under query dependencies.

\section*{Acknowledgments}

We gratefully acknowledge the support of NSERC for grants RGPIN-05849, IRC-537591, and the Royal Bank of Canada for funding this research. 
This work beneﬁted from the use of the CrySP RIPPLE Facility at the University of Waterloo.

\bibliography{mybib}{}
\bibliographystyle{plain}

\appendix
\section{Adapting $\name$ and $\sap$ against $\clrz$ and $\osse$}
\label{app:adapt_osse}

We adapt $\name$ and $\sap$ against $\clrz$ following the approach by Shang et al.~\cite[Appendix D]{shang2021obfuscated}.
Namely, the adversary knows the TPR and FPR of $\clrz$ and computes the $\nkw\times\nkw$ matrix of \emph{expected} keyword volumes $\Vexp$ after the defense is applied.
Recall that $\Vaux_{i,i'}$ is an estimation (from the auxiliary information) of the probability that a document has both keywords $\kw{i}$ and $\kw{i'}$.
Let $\Vauxnot_{i,i'}$ be an estimation of the probability that a document has \emph{neither} keywords $\kw{i}$ nor $\kw{i'}$ (in our experiments, the adversary computes this from the auxiliary data set).
Then, the $i,i'$th entry of $\Vexp$ is~\cite[Appendix D]{shang2021obfuscated}
\begin{equation} \label{eq:Vexp_clrz}
  \Vexp_{i,i'}=\begin{cases} i\neq i':\quad&\TPR^2\cdot \Vaux_{i,i'}+\FPR^2\cdot \Vauxnot_{i,i'}\\
	&+ \TPR\cdot\FPR\cdot(1-\Vaux_{i,i'}-\Vauxnot_{i,i'})\,,\\
	  i=i':\quad&\TPR\cdot\Vaux_{i,i'}+\FPR\cdot\Vauxnot_{i,i'}\,.
		\end{cases}
\end{equation}
To adapt the different attacks against $\clrz$, the adversary simply uses $\Vexp$ instead of $\Vaux$ in the attack's coefficients.

%To adapt the attacks against $\OSSE$, we follow a slightly different approach than Shang et al.~\cite{shang2021obfuscated}.
We explain how we adapt the attacks against $\osse$.
Recall that, in $\osse$, the adversary observes access patterns (with each entry obfuscated with $\TPR$ and $\FPR$), but does not know whether or not two access patterns correspond to the same keyword, since each access pattern has been generated with fresh randomness.
Let $\ntok$ be the number of distinct queried keywords in $\osse$ (the number of distinct observed access patterns could be up to $\nq$; i.e., the number of queries).
Following Shang et al.~\cite{shang2021obfuscated}, the adversary first clusters the observed access patterns into $\ntok$ groups (we assume the adversary knows $\ntok$ for simplicity~\cite{shang2021obfuscated}).
Let $C_j$ be the $j$th cluster, for $j\in[\ntok]$.
To build the off-diagonal $j,j'$th entries ($j\neq j'$) of the matrix of observed volumes $\Vobs$, the adversary computes the \emph{average} number of documents in common between one access pattern from $C_j$ and one from $C_{j'}$, i.e.,
\begin{equation*}
	\Vobs_{j,j'}=\frac{1}{|C_j|\cdot|C_{j'}|}\sum_{\ap{}\in C_j}\sum_{\ap{}' \in C_{j'}} \frac{\ap{}^T\ap{}'}{\ndocs}\,.
\end{equation*}
The diagonal entries are $\Vobs_{j,j}=\frac{1}{|C_j|}\sum_{\ap{}\in C_j} \frac{\ap{}^T\ap{}}{\ndocs}$.
The adversary uses this observation matrix $\Vobs$ and the auxiliary matrix $\Vexp$ above \eqref{eq:Vexp_clrz} to run the attacks against $\osse$.
\section{Adapting $\name$ against $\pancake$}
\label{app:adapt_pancake}

We provide details into how we adapt $\name$ against $\pancake$.
As we mention in the main text, the client chooses keywords for her queries following a Markov process characterized by the $\nkw\times\nkw$ matrix $\Freal$, with stationary profile $\freal$.
Every time the client queries the server, $\pancake$ creates three query slots, selects three keywords, and sends the corresponding query tokens to the server (see Section~\ref{sec:pancake}).
Each of these keywords can either be \emph{dummy} ($\kw{}\sim\fdum$), \emph{real fake} ($\kw{}\sim\freal$), or an actual real query (sampled from $\Freal$, according to the previous real query).
Let $T_1,T_2,\dots,T_\nq$ be the set of query token triplets observed by the adversary.
The adversary builds the matrix of observed token frequencies $\Fobs$ by counting all token transitions in consectuve triplets (Algorithm~\ref{alg:Fobs}).
\begin{algorithm}[t]
    \caption{Compute $\Fobs$ in $\pancake$}
    \label{alg:Fobs}
    \begin{algorithmic}[1] % The number tells where the line numbering should start
        \Procedure{ComputeFreqObs}{$\nkw$,$T_1,T_2,\dots,T_\nq$} %\Comment{The g.c.d. of a and b}
				\State $\Fobs\gets \mathbf{0}_{\nrep\times\nrep}$
					\For{$i\gets 1 \dots \nq-1$}
						\For{$\tok{j}\in T_i$, $\tok{j'}\in T_{i+1}$}
							\State $\Fobs_{j',j}\gets \Fobs_{j',j}+1$
						\EndFor
					\EndFor
				\State Normalize columns of $\Fobs$.
				\State \textbf{return} $\Fobs$
        \EndProcedure
    \end{algorithmic}
\end{algorithm}

\begin{figure*}[t]
\begin{minipage}{0.20\linewidth}
\begin{center}
\def\svgwidth{\linewidth} 
{
%\fontsize{7.5pt}{9.5pt}
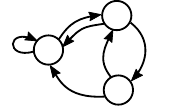
}
\end{center}
\end{minipage} \hfill
\begin{minipage}{0.40\linewidth}
\begin{center}
\def\svgwidth{0.9\linewidth} 
{
%\fontsize{7.5pt}{9.5pt}
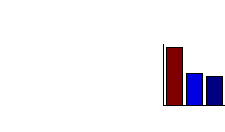
}
\end{center}
\end{minipage} \hfill
\begin{minipage}{0.33\linewidth}
\begin{center}
\def\svgwidth{0.9\linewidth} 
{
%\fontsize{7.5pt}{9.5pt}
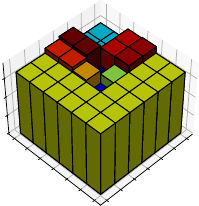
}
\end{center}
\vspace{5pt}
\end{minipage}\\
\begin{minipage}{0.20\linewidth}
\centering
(a) Markov model
\,\\
\end{minipage} \hfill
\begin{minipage}{0.40\linewidth}
\centering
(b) Markov matrix ($\Freal$) and its stationary distribution ($\freal$) of the queried keywords.
\end{minipage} \hfill
\begin{minipage}{0.33\linewidth}
\centering
(c) Markov matrix ($\Fexp$) of the queried replicas by following $\pancake$ protocol.
\end{minipage}
\caption{Example of $\pancake$ under correlated queries. \label{fig:pancake_app}}
\end{figure*}

Then, the adversary builds the $\nrep\times\nrep$ expected matrix of \emph{replica} frequencies $\Fexp$ given the auxiliary information $\Faux$.
First, the adversary computes $\faux$ from $\Faux$, and gets the number of replicas of each keyword and the dummy keyword distribution $\fdumaux$ following $\pancake$'s specifications.
Let $\tilde{R}(i)$ be the number of replicas for keyword $\kw{i}$ and $\reptokw: [\nrep]\to[\nkw]$ be a mapping from replicas to keywords, both computed from $\faux$.

If we generate keyword queries following $\pancake$'s specifications and using $\Faux$, $\faux$ and $\fdumaux$, for every two keywords $\kw{}$ and $\kw{}'$ in consecutive triplets, one of the following events happens:
\begin{enumerate}[$A$.]
	\item $\kw{}'$ was sampled from $\fdumaux$.
	\item $\kw{}$ was sampled from $\fdumaux$ but $\kw{}'$ was not.
	\item Neither $\kw{}$ nor $\kw{}'$ were sampled from $\fdumaux$, but at least one is a ``real fake'' query sampled from $\faux$.
	\item Both $\kw{}$ and $\kw{}'$ are real queries.
\end{enumerate}

	%\noindent-$A$: $\kw{}'$ was sampled from $\fdumaux$.\\
	%-$B$: $\kw{}$ was sampled from $\fdumaux$ but $\kw{}'$ was not.\\
	%-$C$: Neither $\kw{}$ nor $\kw{}'$ were sampled from $\fdumaux$, but at least one is a fake query sampled from $\faux$.\\
	%-$D$: Both $\kw{}$ and $\kw{}'$ are real queries.

Following $\pancake$ specifications, one can see that $\Pr(A)=0.5$ and $\Pr(B)=0.25$.
We computed the probabilities of events $C$ and $D$ empirically, since they are constants that just depend on $\pancake$'s specifications: $\Pr(C)\approx 0.145$ and $\Pr(D)\approx 0.105$.
Note that $\kw{}$ and $\kw{}'$ are independent except in event $D$.
If both $\kw{}$ and $\kw{}'$ are keywords for real queries (event $D$), then $\kw{}'$ was generated right after $\kw{}$ with probability 0.81, two queries after $\kw{}$ with probability $0.17$, and three queries after $\kw{}$ with probability $0.02$ (we also determined these probabilities empirically; they are constants that only depend on $\pancake$'s protocol).
Summarizing, with probability 0.5 we have that $\kw{}'\sim\fdumaux$, with probability $0.25+0.145=0.395$ we have $\kw{}'\sim\faux$, and otherwise $\kw{}'$ depends on $\kw{}$.
Putting this together, the transition probabilities between two keywords in consecutive triplets are:
\begin{equation*}
\resizebox{\hsize}{!}{
	$\mathbf{G}=0.105 \cdot (0.81\Faux+0.17\Faux^2+0.02\Faux^3) + 0.395\cdot \faux\mathbf{1}_\nkw^T+0.5\cdot\fdumaux\mathbf{1}_\nkw^T\,.$
}
\end{equation*}
Matrix $\Fexp$ follows this formula, expanded to the space of replicas:
\begin{equation} \label{eq:Fexp}
 \Fexp_{j,j'}=\mathbf{G}_{\reptokw(j),\reptokw(j')}/\tilde{R}(\reptokw(j))\,.
\end{equation}

Figure~\ref{fig:pancake_app} shows an example of a Markov model with three keywords (Fig.~\ref{fig:pancake_app}a), the Markov matrix and stationary profile (Fig.~\ref{fig:pancake_app}b), and the expected transition between replicas $\Fexp$ \eqref{eq:Fexp} (Fig.~\ref{fig:pancake_app}c).
In the plot, we used $\Faux=\Freal$ to compute $\Fexp$, for illustration purposes.
The stationary profile of $\Fexp$ is uniform, but we can see that $\Fexp$ is not.
The matrix of observed frequencies $\Fobs$ is a (noisy) version of $\Fexp$, with randomly permuted rows and columns.
We use $\name$ to estimate this permutation by setting the coefficients of $\name$ as in \eqref{eq:coefFreq1} and \eqref{eq:coefFreq2}, but using $\Fexp$ instead of $\Faux$.
The result of this is a mapping $\P$ from tokens to replicas, which we map to keywords using $\reptokw(\cdot)$.

\section{Dataset Generation}
\label{app:dataset}

Table~\ref{tab:datasets} shows the dataset names, the type and number of documents they contains, and the source URL that we used to download them.
We download the datasets from the source we show in the table and parse them so that each document is a string (an email, news article, or movie plot summary).
Following related work~\cite{damie2021highly}, for Enron dataset we only take documents in the \texttt{\_sent\_mail} folder; for Lucene, we remove the signature at the end of each email that begins with \texttt{``To unsubscribe''}, since this message appears in all emails.
Then, we follow a series of steps to extract the keywords of each document:
\begin{enumerate}
	\item We extract each word in the document using the regular expression (regex) \verb|\w+|.
	\item We convert all words to lower-case and ignore words that contain non-alpha characters~\cite{poddar2020practical}.
	\item We ignore English stopwords, and words whose length is not between 3 and 20 characters~\cite{poddar2020practical}.
	\item We apply Porter stemmer~\cite{van1980new} to the remaining words, and keep the $3\,000$ most popular stems of each dataset.
	This stemming process is the most popular approach to extract keywords in related work~\cite{islam2012access, cash2015leakage, blackstone2020revisiting, poddar2020practical, damie2021highly, ning2021leap}.
\end{enumerate}
These stems are the keywords in our evaluation.
We save, for each dataset, the different words that yielded each of the $3\,000$ stems.
Then, for each of those words, we download their query frequencies for each week of 2020 from Google Trends.\footnote{\url{https://trends.google.com/trends}}
We use the \texttt{gtab} package~\cite{west2020calibration} to fix normalization issues of Google Trends.
The query frequency of a particular stem is the sum of frequencies for each of its keywords.
For example, \texttt{``time''} is a popular stem in Enron dataset.
The words that resulted in this stem are ``time'', ``timed'', ``timely'', etc.
We downloaded the query frequencies for each of those words, and assumed that queries for each of those words trigger a match in documents that contain the stem \texttt{``time''}.

\begin{figure}[t]
\centering
\includegraphics[width=\linewidth]{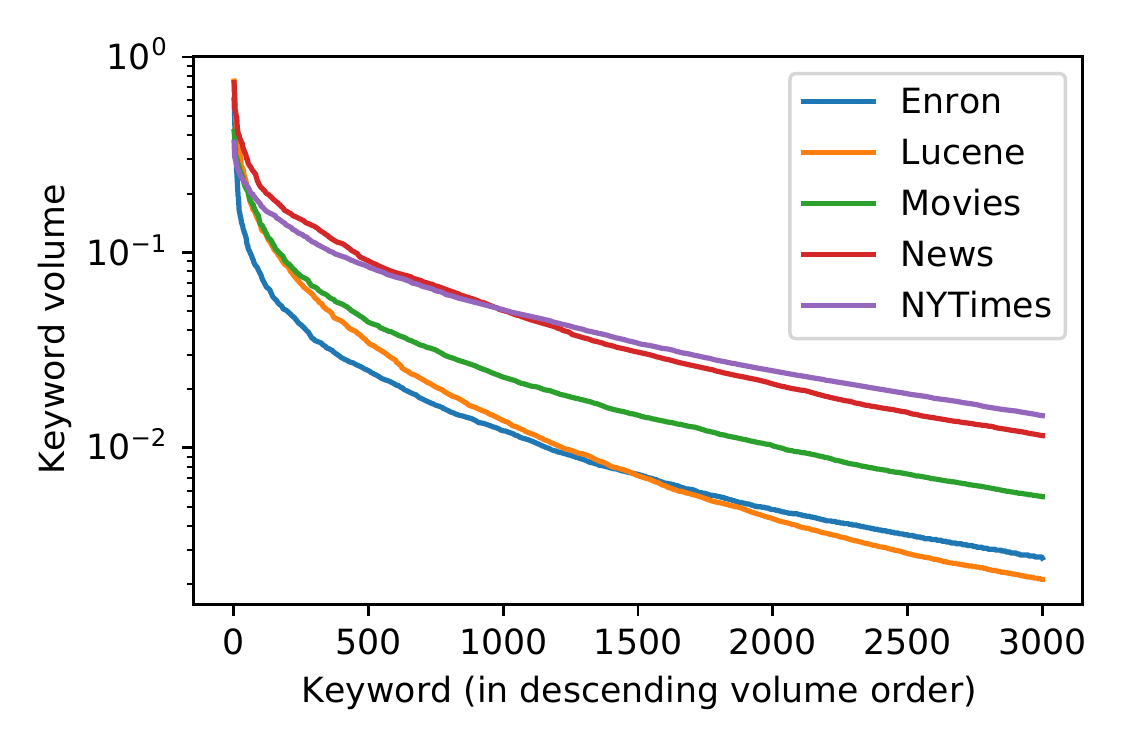}
\caption{Keyword volume distribution for each of the datasets we consider.}
\label{fig:datasets}
\end{figure}

Figure~\ref{fig:datasets} shows the volume of each keyword in the datasets.
Recall that the volume of a keyword is the percentage of documents that contain such keyword.
We sort the keywords of each dataset in decreasing volume order in the plot.
In email datasets (Enron and Lucene), we see that keywords have lower volumes than in other datasets, which indicates that each email uses specific terminology that is not very common among other emails.
Datasets that consist of news articles (News and NYTimes) lie in the opposite extreme: their documents use a similar vocabulary, which yields keywords with overall high volumes.
The volume distribution of the dataset that consists of movie summaries (Movies) is somewhere in between these two extremes.
This plot shows that the datasets we consider are statistically varied.
We note that the volume distribution is not necessarily correlated with attack accuracy.
There are many variables that affect attack accuracy (e.g., keyword volume uniqueness, keyword co-occurrence uniqueness, higher-order dependencies between keywords), and we cannot capture all of them in a single plot.

\begin{table*}[t]
\centering
\begin{tabular}{ r  c l l }
  Dataset & Size & Content & Source \\ \hline
	Enron & $30\,033$ & Emails & \url{https://www.cs.cmu.edu/~./enron/} \\
	Lucene & $63\,597$ & Emails & \url{https://mail-archives.apache.org/mod_mbox/lucene-java-user/} \\
	Movies & $42\,304$ & Movie plots & \url{http://www.cs.cmu.edu/~ark/personas/} \\
	News & $49\,931$ & Articles & \url{https://www.kaggle.com/snapcrack/all-the-news/version/4} \\
	NYTimes & $299\,607$  & Articles & \url{https://archive.ics.uci.edu/ml/datasets/bag+of+words}
\end{tabular}
\caption{Datasets that we consider. \label{tab:datasets}}
\end{table*}
\section{Wikipedia Dataset Generation}
\label{app:wiki}

We explain how we generate the keyword universes and transition matrices $\Freal$ for our experiments with query correlations (Section~\ref{sec:eval_pancake}).
As we mention in the main text, we consider a scenario where each keyword-document pair represents a particular topic or webpage, and we use the Wikipedia clickstream dataset to build a realistic transition matrix $\Freal$ between keywords.
Thus, we refer to keywords as \emph{pages} in this appendix.

We use PetScan\footnote{\url{https://petscan.wmflabs.org}} to retrieve all Wikipedia pages under the categories  \emph{`privacy'}, \emph{`security'}, \emph{`cryptography'}, \emph{`politics'}, and \emph{`activism'}, including pages in subcategories up to a depth of two.
This yields $5\,573$ pages.
We download the number of times Wikipedia users transitioned between those pages in 2020 by querying the Wikipedia Clickstream dataset.\footnote{\url{https://dumps.wikimedia.org/other/clickstream/}}
Let $G_{big}$ be a graph where the nodes are these pages, and the edges between two nodes are the number of times users transitioned between the pages represented by such nodes.

We explain how we build a keyword universe $\kwu$ of size $\nkw=500$ centered around a category.
We start with a subgraph $G_{cat}\subset G_{big}$ with nodes (pages) from that category only.
We remove all nodes with a degree $\leq 1$ and we keep removing nodes with the smallest degree in $G_{cat}$ until the subgraph has a size smaller than $\nkw$.
If the subgraph already had less nodes than $\nkw$, we add nodes from $G_{big}$ (the ones that share more edges with $G_{cat}$ are added first) until $G_{cat}$ has $\nkw$ nodes.
Then, let $\kw{big}$ be the node in $G_{big}\setminus G_{cat}$ that has more edges connecting to $G_{cat}$ (and let $s(\kw{big})$ be this number of edges), and let $\kw{cat}$ be the node in $G_{cat}$ with smallest degree (and let $s(\kw{cat})$ be this degree).
If $s(\kw{cat})\leq 1$ or if $s(\kw{cat})<s(\kw{big})-2$, we remove $\kw{cat}$ from $G_{cat}$ and replace it with $\kw{big}$.
Otherwise, we finish the building process, and the remaining $\nkw$ nodes in $G_{cat}$ are the keyword universe $\kwu$ for the category.

Finally, we build $\Fnew$ from $\kwu$ as follows (the process for $\Fold$ is the same, with data from different months).
We get all transitions between pages in $\kwu$ from July to December (2020), and also get the number of times users accessed pages in $\kwu$ from \emph{other} sources (namely, transitions from pages named \textit{`other-empty'}, \textit{`other-external'}, \textit{`other-internal'}, \textit{`other-search'}, \textit{`other-other'} in the Clickstream dataset).
We use these transitions from other soruces to compute the query probability of the pages in $\kwu$ when the user starts a browsing session (we denote these probabilities by $\mathbf{p}_{other}$).

The columns of $\Fnew$ are transition profiles, i.e., the $i$th column is a vector that represents the probability of querying for a particular page after querying for $\kw{i}$.
To compute this profile for a page $\kw{i}\in\kwu$, we count the transitions from $\kw{i}$ to all other $\kw{i'}\in\kwu$ and normalize this so that it adds up to one.
We also consider that there is a probability $0.05$ of \emph{restarting} the browsing session, i.e., with probability $0.05$ the user simply chooses a new page by following $\mathbf{p}_{other}$.
This ensures the Markov process is irreducible and aperiodic.
If $\kw{i}$ is a \emph{sink page}, i.e., a page the user can transition to but that does not have outgoing transitions to other pages in $\kwu$, we simply consider that the user restarts the browsing session after that page, so the transition profile is just $\mathbf{p}_{other}$.

\section{Evaluation Results in All Datasets}
\label{app:exp}

\begin{figure*}[b]
\centering
\includegraphics[width=0.95\linewidth]{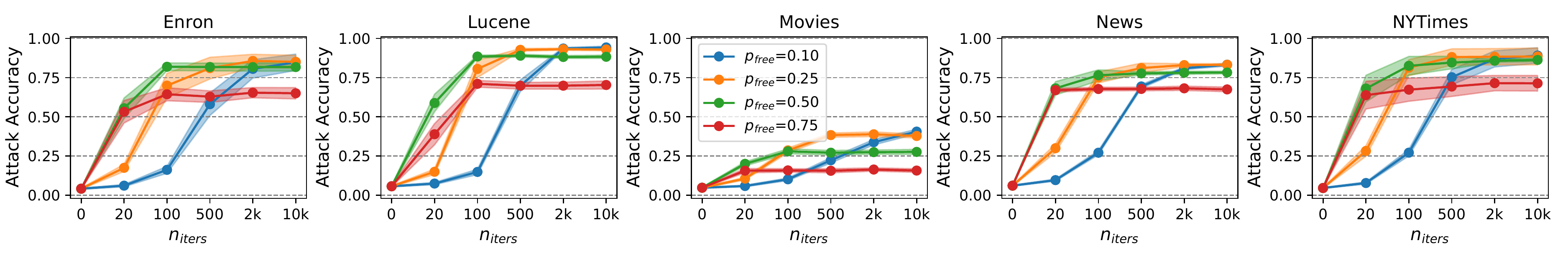}
\vspace{-0.2cm}
\caption{Evolution of $\name$'s accuracy with the number of iterations and different $\pfree$ values. (S1)}
\vspace{0.3cm}
\label{fig:exp0_all}
\includegraphics[width=0.95\linewidth]{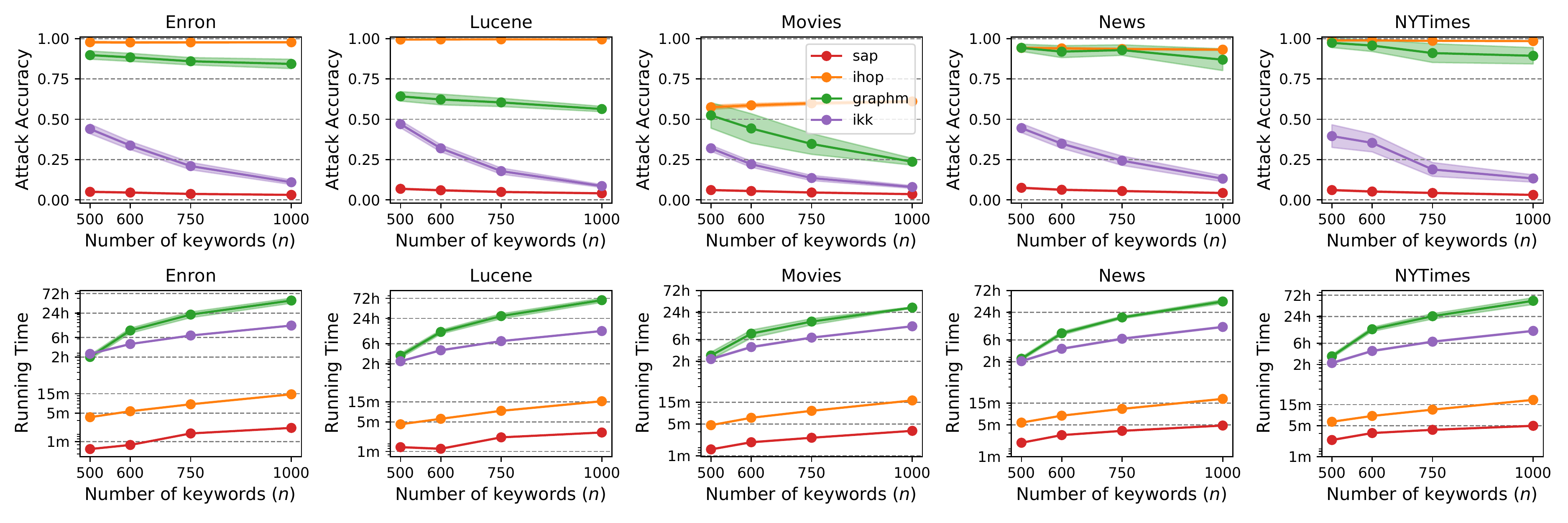}
\vspace{-0.2cm}
\caption{Attack comparison vs.~number of keywords ($n$) (S1)}
\vspace{0.3cm}
\label{fig:exp11_all}
\includegraphics[width=0.95\linewidth]{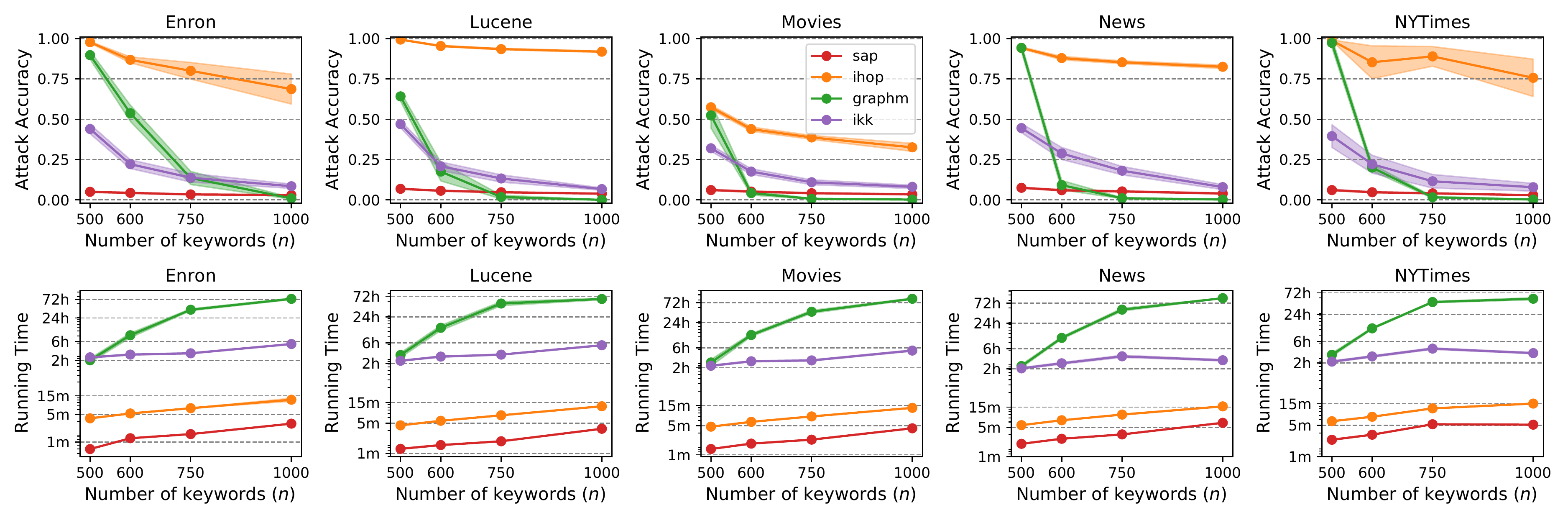}
\vspace{-0.2cm}
\caption{Attack comparison vs.~number of keywords ($n$) when the adversary sees $\ntok=500$ distinct tokens (S2)}
\vspace{0.3cm}
\label{fig:exp11_fixed_all}
\includegraphics[width=0.95\linewidth]{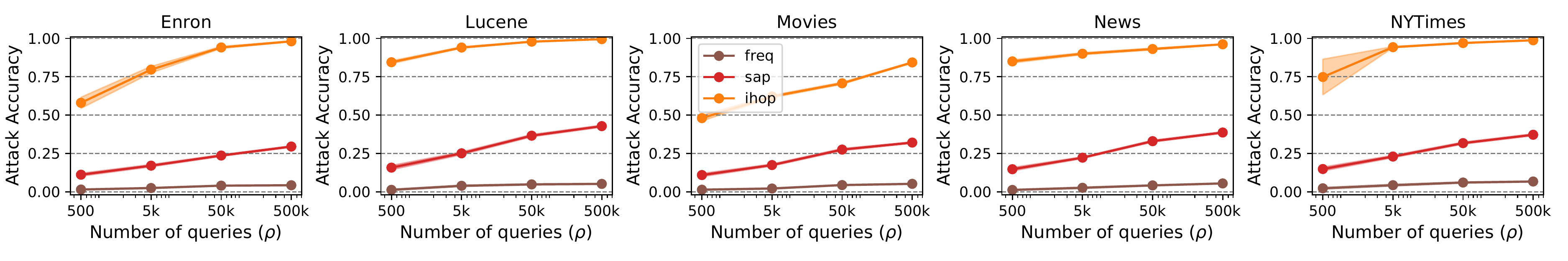}
\vspace{-0.2cm}
\caption{Comparison of attacks that use frequency leakage, with $\nkw=3\,000$ (S2)}
\vspace{0.3cm}
\label{fig:s2_all}
\end{figure*}

\begin{figure*}
\centering
\includegraphics[width=0.95\linewidth]{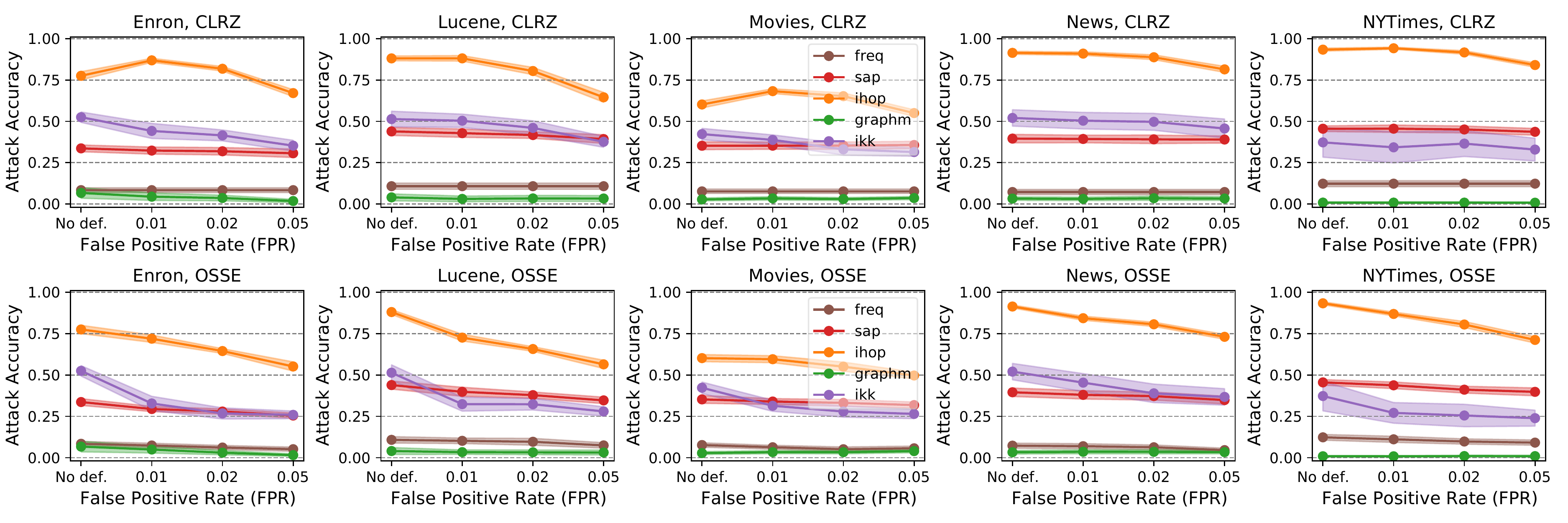}
\caption{Attack comparison vs.~$\clrz$ and $\osse$ ($\nkw=500$, $\nq=500$) (S2)}
\label{fig:s21_all}
\end{figure*}

For completeness, below we show the results of all our experiments in the five datasets.
Figures \ref{fig:exp0_all}, \ref{fig:exp11_all}, \ref{fig:exp11_fixed_all}, \ref{fig:s2_all}, and \ref{fig:s21_all} below correspond to the experiments shown in Figures~\ref{fig:exp0}, \ref{fig:exp11}, \ref{fig:exp11_fixed}, \ref{fig:s2}, and \ref{fig:s21}, respectively. 
In all these figures, we can see that the results are qualitatively similar: the trends of $\name$ with $\niters$ and $\pfree$ are the same for all datasets in Figure~\ref{fig:exp0_all}, and $\name$ outperforms all other attacks in the remaining experiments.

\end{document}